\documentclass[floats,floatfix,showpacs,amssymb,prd,twocolumn,superscriptaddress,nofootinbib,nolongbibliography,reprint]{revtex4-1}

\usepackage{amssymb,amsmath,verbatim,mathtools,needspace,enumitem,etoolbox,graphicx,physics,microtype,afterpage}
\usepackage{gensymb}

\usepackage[dvipsnames, usenames]{xcolor}

\definecolor{linkcolor}{rgb}{0.0,0.3,0.5}
\usepackage[unicode, colorlinks=true, linkcolor=linkcolor, citecolor=linkcolor, filecolor=linkcolor,urlcolor=linkcolor, pdfusetitle]{hyperref}
\usepackage[all]{hypcap}
\usepackage[T1]{fontenc}
\usepackage[utf8]{inputenc}
\usepackage{tabularx}
\usepackage{lmodern}

\newcommand{\jhu}{\affiliation{Department of Physics and Astronomy, Johns Hopkins University, 3400 N. Charles Street, Baltimore, Maryland 21218, USA}}
\newcommand{\bham}{\affiliation{School of Physics and Astronomy \& Institute for Gravitational Wave Astronomy, \\ University of Birmingham, Birmingham, B15 2TT, UK}}

\def\pa{\partial}

\def\nn{\nonumber}

\newcommand{\ben}{\begin{enumerate}}
\newcommand{\een}{\end{enumerate}}

\def\be{\begin{equation}}
\def\ee{\end{equation}}
\def\bea{\begin{eqnarray}}
\def\eea{\end{eqnarray}}
\def\nn{\nonumber}
\newcommand{\beq}{\begin{eqnarray}}
\newcommand{\eeq}{\end{eqnarray}} 
\newcommand{\ba}{\begin{align}}
\newcommand{\ea}{\end{align}}

\definecolor{royalblue4}{HTML}{27408B}

\newcommand{\new}[1]{{{#1}}}

\def\nn{\nonumber}
\def\be{\begin{equation}}
\def\ee{\end{equation}}
\def\beq{\begin{eqnarray}}
\def\eeq{\end{eqnarray}}
\def\f{\frac}

\newcommand{\mbh}{\langle m_{\rm BH}\rangle}
\newcommand{\mstar}{\langle m_{\star}\rangle}
\newcommand{\Mcl}{M_{\mathrm{cl}}}
\newcommand{\Mbh}{M_{\rm BH}}
\newcommand{\Msun}{\,{\rm M}_{\odot}}

\newcommand{\nstar}{n_{\star}}
\newcommand{\nbh}{n_{\rm BH}}
\newcommand{\sigbh}{\sigma_{{\rm BH}}}
\newcommand{\sigstar}{\sigma_{\star}}
\newcommand{\aej}{a_{\rm ej}}
\newcommand{\agw}{a_{\rm GW}}
\newcommand{\tthreebb}{t_{\mathrm{3bb}}}
\newcommand{\tonetwo}{t_{1\text{-}2}}
\newcommand{\tgw}{t_{\mathrm{GW}}}
\newcommand{\ttot}{t_{\mathrm{tot}}}

\newcommand{\kms}{\mathrm{km\,s^{-1}}}
\newcommand{\acrit}{a_{\mathrm{crit}}}
\newcommand{\ahard}{a_{\mathrm{hard}}}

\newcommand{\tms}{t_{\mathrm{MS}}}
\newcommand{\thard}{t_{\mathrm{hard}}}
\newcommand{\rh}{r_{\rm h}}
\newcommand{\rhobh}{\rho_{{\rm BH}}}
\newcommand{\pcc}{\mathrm{pc}^{-3}}
\newcommand{\vkick}{v_\mathrm{kick}}
\newcommand{\vesc}{v_\mathrm{esc}}
\newcommand{\tlb}{t_{\rm lookback}}

\newcommand{\chimax}{\chi_{\rm max}}
\newcommand{\chieff}{\chi_{\rm eff}}
\newcommand{\mmin}{m_{\rm min}}
\newcommand{\mmax}{m_{\rm max}}

\newcommand{\chihat}{\hat{\chi}}
\newcommand{\phat}{\hat{p}}
\newcommand{\pcl}{\hat{p}_{\rm cluster}}
\newcommand{\pf}{\hat{p}_{\rm field}}

\newcommand{\lspin}{\lambda_{\rm S}}
\newcommand{\lmass}{\lambda_{\rm M}}
\newcommand{\Ncl}{N_{\rm cluster}}
\newcommand{\Nf}{N_{\rm field}}
\newcommand{\Nspin}{N_{\rm Sgap}}
\newcommand{\Nmass}{N_{\rm Mgap}}
\newcommand{\sL}{{\mathcal L}}
\newcommand{\chierr}{{\overline{\delta\chieff}}}

\def\nn{\nonumber}
\def\be{\begin{equation}}
\def\ee{\end{equation}}
\def\beq{\begin{eqnarray}}
\def\eeq{\end{eqnarray}}
\def\f{\frac}

\def\apjl{\rm{ApJ}}
\def\apjs{\rm{ApJS}}
\def\aap{\rm{A\&A}}

\def\mnras{\rm{MNRAS}}

\begin{document}

\title{The mass gap, the spin gap, and the origin of merging binary black holes}

\author{Vishal Baibhav}
\email{vbaibha1@jhu.edu}
\jhu
\author{Davide Gerosa}
\bham
\author{Emanuele Berti}
\jhu
\author{Kaze W. K. Wong}
\jhu
\author{Thomas Helfer}
\jhu
\author{Matthew Mould}
\bham
\pacs{}
\date{\today}

\begin{abstract}
  Two of the dominant channels to produce the black-hole binary mergers observed by LIGO and Virgo are believed to be the isolated evolution of stellar binaries in the field and dynamical formation in star clusters. Their relative efficiency can be characterized by a ``mixing fraction.'' Pair instabilities prevent stellar collapse from generating black holes more massive than about $45 \Msun$. This ``mass gap'' only applies to the field formation scenario, and it can be filled by repeated mergers in clusters. A similar reasoning applies to the binary's effective spin.
If black holes are born slowly rotating, the high-spin portion of the parameter space (the ``spin gap'')  can only be populated by black hole binaries that were assembled dynamically. 
Using a semianalytical cluster model, we show that future gravitational-wave events in either the mass gap, the spin gap, or both can be leveraged to infer the mixing fraction between the field and cluster formation channels.
\end{abstract}

\maketitle

\section{Introduction}

Gravitational-wave (GW) observations of merging black-hole (BH) binaries are bringing us into a new era where many questions are still unanswered. How, when, and where do these binaries form? What is the core physics that drives them to merge? 

The two most popular formation channels are isolated binary evolution in the field and dynamical formation in clusters (see e.g.~\cite{Mandel:2018hfr,Mapelli:2018uds} for reviews).  For isolated binaries, the most promising mechanism to catalyze mergers is a common-envelope phase in between the formation of the two BHs. Alternatively, dynamical channels predict that binary BHs (BBHs) form and harden through three-body encounters in dense stellar clusters. \new{Other scenarios for the formation and merger of BBHs include chemically homogenous evolution~\cite{Marchant:2016wow,deMink:2016vkw}, AGN disks~\cite{Leigh:2017wff,Stone:2016wzz,Bartos:2016dgn}, secular interactions in triples~\cite{Silsbee:2016djf,Hoang:2017fvh,Fragione:2018yrb}, and primordial BHs~\cite{Bird:2016dcv}.} Different formation pathways leave different imprints on the properties of the BBH population, including the binary masses, spins, eccentricities, and redshift evolution. Measuring these distributions informs us on the environment in which BBHs form and evolve~\cite{Zevin:2017evb,Taylor:2018iat,Wysocki:2018mpo, Roulet:2018jbe,LIGOScientific:2018jsj}.

One of the most promising signatures is the distribution of BH spins: systems formed through dynamical interactions are expected to have isotropic spin orientations, whereas binaries born in the field are more likely to have aligned spins~\cite{Gerosa:2013laa,Rodriguez:2016vmx,Farr:2017gtv,Gerosa:2018wbw}.

However, if BHs are naturally born with low spins, it becomes harder to differentiate between formation channels using spin alignment.  LIGO/Virgo observations indicate that this may be the case: the majority of the events reported so far have ``effective spin''\footnote{For a binary with component masses $m_1>m_2$, mass ratio $q=m_2/m_1$ and dimensionless spins of magnitude $\chi_i$ at angles $\theta_i$ ($i=1,\,2$) with respect to the orbital angular momentum, the effective spin $\chieff\equiv (\chi_1 \cos\theta_1+q \chi_2\cos\theta_2)/( 1+q)$ is a mass-weighted combination of the components of the BH spins parallel to the binary's orbital angular momentum.} $\chieff \simeq0$. More specifically, all detections from Ref.~\cite{LIGOScientific:2018mvr} but two (GW151226 and GW170729) are compatible with $\chieff=0$ at 90\% confidence, although this is a somewhat prior-dependent statement~\cite{Vitale:2017cfs}. For GW151226 and GW170729, the 90\% lower limit on the effective spin is as low as $\chieff \sim 0.1$~\cite{LIGOScientific:2018mvr}.  A recent study \cite{Miller:2020zox} showed that the effective spin distribution of LIGO observations is almost consistent with a Dirac delta centered at $\chieff=0$.  The additional triggers reported in Refs.~\cite{Zackay:2019tzo,Zackay:2019btq}, if astrophysical in nature, might be high-spin outliers with $\chi_{\rm eff}\gtrsim 0.5$, but this is also a prior-dependent statement~\cite{Huang:2020ysn}.

Recent stellar-physics simulations also suggest that BHs are born with very low spins. Efficient core-envelope interactions may transfer the angular momentum of the progenitor star away from the collapsing core, resulting in BH spins $\chi\sim10^{-2}$~\cite{Fuller:2019sxi}.

If stellar-mass BHs do indeed rotate very slowly, we will not be able to differentiate the aligned and isotropic populations, making it difficult (if not impossible) to use spin alignment to disentangle BH mergers formed in the field from \new{dynamically-formed binaries}.

In this paper we identify specific observational signatures that are enhanced if spins are indeed small. In a nutshell, we exploit specific regions of the parameter space which can plausibly be populated by only one of the two scenarios. These ``reserved regions'' or ``gaps'' provide a new handle to infer the mixing fraction between the underlying formation channels.

{Suppose, for simplicity, that only two formation channels (``field'' and ``cluster'') are at play for $N$ BBH detections: 
\begin{equation}
\Nf + \Ncl = N\,.
\end{equation}
The fraction of observation from the ``cluster'' scenario is
\be
f\equiv \f{\Ncl}{N}\,.
\ee
while the ``field'' fraction is given by $1-f$.
Let us further separate the fraction of the catalog entries that are inside/outside a specific region of the parameter space (``gap''), i.e.
\begin{align}
N &= N_{\rm no\,gap} + N_{\rm gap}\,.
\end{align}
 This gap is a reserved region, in the sense that it can only be populated by one of the models (say ``cluster''): this implies $N_{\rm field,gap}=0$, and therefore $N_{\rm gap}= N_{\rm cluster,gap}$.
If the efficiency of the ``cluster'' model at populating the gap
\begin{equation}
\lambda \equiv  \frac{N_{\rm cluster,gap}}{N_{\rm cluster}}
\end{equation}
can be reliably  estimated, one immediately obtains an estimate of the number of binaries coming from each population:
\be
\Ncl = \f{N_{\rm gap}}{\lambda}\,,\quad\quad\quad \Nf = N-\f{N_{\rm gap}}{\lambda}\,,
\ee
or equivalently of the mixing fraction: 
\be
f= \frac{N_{\rm gap}}{\lambda\ N}\,.
\ee
For instance, if $N\sim 100$ events are detected during LIGO/Virgo's third observing run O3 and one of them lies in the gap, an efficiency $\lambda\sim 5\%$ would imply that $f\sim 20\%$ of the observed BH binaries must have formed in clusters, and the remaining $1-f\sim 80\%$ must have formed in the field.}

\begin{figure}[t]
  \includegraphics[width=\columnwidth]{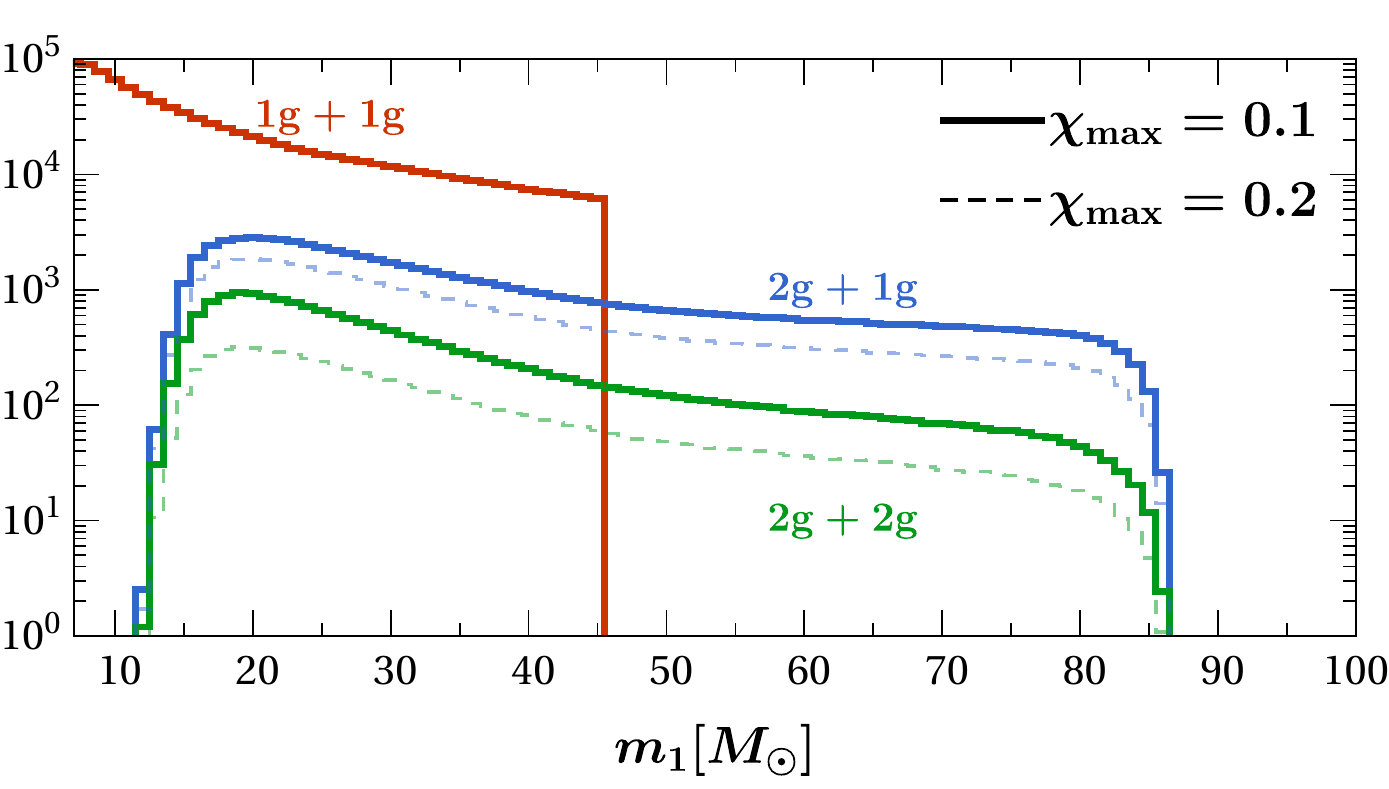}
  \includegraphics[width=\columnwidth]{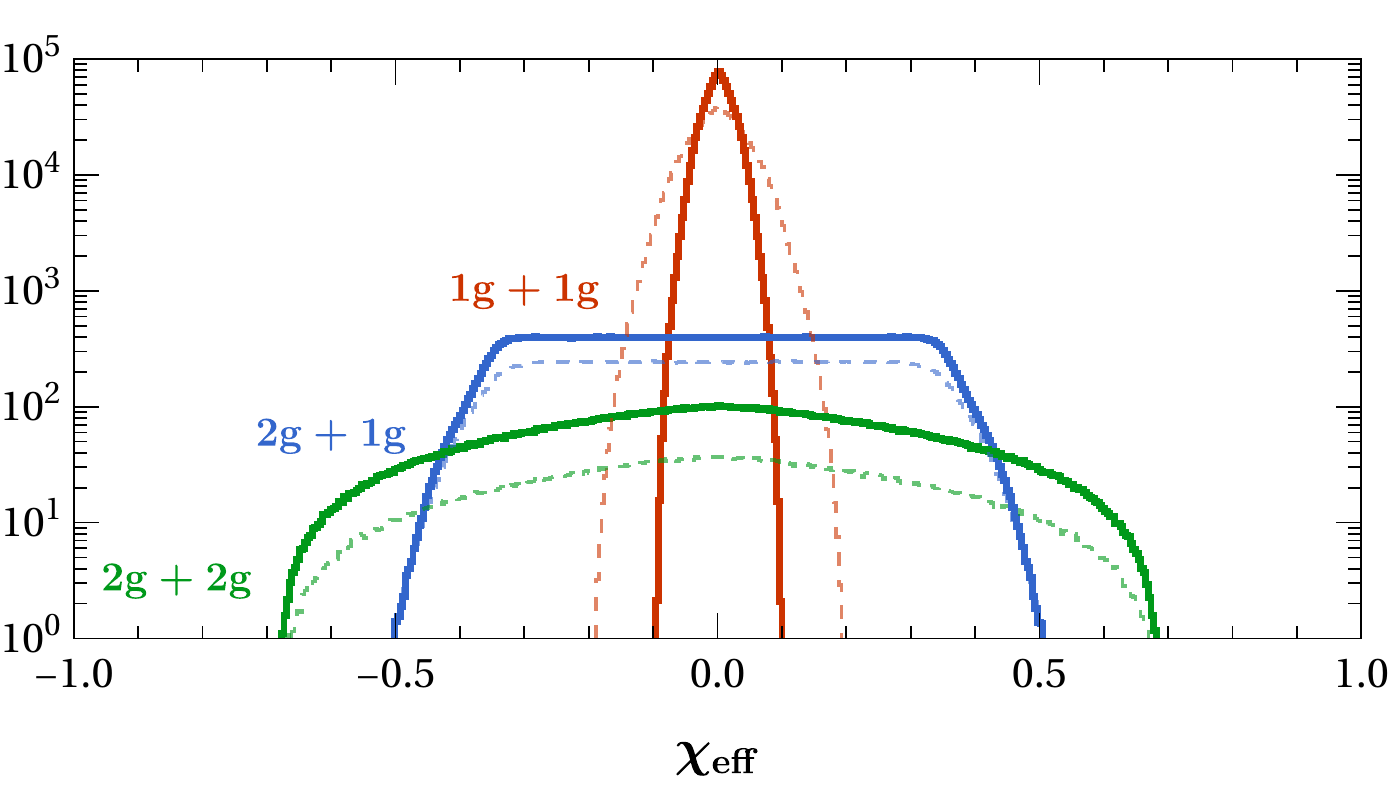}
  \caption{Illustration of the ``mass gap'' in    the primary mass $m_1$ (top panel) and of the ``spin gap'' in the effective spin $\chieff$ (bottom panel). Solid (dashed) lines are computed under the assumption that the maximum individual BH spin at birth is $\chimax=0.1\,(0.2)$. Only 2g events can populate the regions of the parameter space with high values of $m_1$ and/or $\chi_{\rm eff}$.}\label{fig:hist}
\end{figure}

Gaps in the parameter space are naturally populated by hierarchical BH mergers. When two BHs merge in the field, the remnant BH does not interact again with other BHs. This is not necessarily true for BHs that merge in clusters. If these ``second-generation'' (henceforth 2g) BHs remain in the cluster, they might continue to interact with other BHs, eventually forming new binaries and merging again \cite{Gerosa:2017kvu,Fishbach:2017dwv,Rodriguez:2019huv,Gerosa:2019zmo}.  These 2g BHs will, on average, be heavier than their ancestors. Moreover, binary formation and hardening tend to occur faster for heavier objects, and thus mergers occur more often.

Both supernova models and LIGO observations \cite{LIGOScientific:2018jsj} indicate the presence of a mass gap (usually referred to as the ``upper mass gap'', to distinguish it from the putative gap between BHs and neutron stars). Pair-instability supernova (PISN) and pulsational pair-instability supernova (PPISN)~\cite{Woosley:2016hmi} prevent the formation of BHs with masses larger than $\sim 45 M_\odot$~\cite{Belczynski:2016jno,Renzo:2020rzx,Farmer:2019jed,DiCarlo:2019fcq,Stevenson:2019rcw}. The pair-instability mass gap is our first reserved region: if a merging binary with a component BH heavier than the PISN threshold is found, this would point towards a hierarchical origin.   

When two ``first-generation'' (henceforth 1g) BHs merge, they form a remnant with a unique distribution of spins which is largely independent of the spins of their progenitors. In particular, remnant spins are strongly peaked at $\chi\sim0.7$~\cite{Berti:2008af,Gerosa:2017kvu,Fishbach:2017dwv}. This is our second reserved region, which we call the ``spin gap'' (although to be rigorous we should call it the ``effective spin gap''): if BHs are indeed born with low spins from stellar collapse, the detection of a highly spinning object would also indicate a hierarchical origin. The mass and spin gaps are illustrated in Fig.~\ref{fig:hist}. 

A 2g merger can occur only if (i) the preceding 1g merger happened in situ, and (ii) the merger remnant remains bound to the cluster. Only BHs that receive kicks smaller than the escape speed of the clusters can be retained and  potentially merge again.  Conversely, the detection of 2g mergers can be used to constrain the escape speed of clusters~\cite{Gerosa:2019zmo,Belczynski:2020bnq}. Generic BH recoils are $\mathcal{O}(100\,{\rm km/s})$~\cite{Gerosa:2019zmo,Gerosa:2018qay}, but kick velocities tend to zero for BHs with similar masses and small spins, as indicated by current observations.

\new{While we assume that 2g mergers happen only in dense star cluster, other astrophysical mechanisms (such as gas accretion~\cite{Roupas:2019dgx}, stellar mergers~\cite{DiCarlo:2019fcq}, Population III stars~\cite{Kinugawa:2015nla,Hartwig:2016nde,Belczynski:2016ieo} or gravitational lensing~\cite{Ng:2017yiu,Hannuksela:2019kle}) could lead to events that contaminate these gaps and complicate the measurement of the mixing fraction, $f$. However these mechanisms are expected to be subdominant. Furthermore it should still be possible to disentangle the population of dynamically formed 2g mergers from other sources, because of the unique relationship between the 1g and 2g populations.}

The rest of this paper puts these ideas on more solid footing. In Sec.~\ref{sec:clusters} we describe a semianalytical cluster model based on simple prescriptions, which, however, can replicate \new{the main features relevant to BH mergers of the more complex and computationally expensive Monte Carlo simulations~\cite{Kremer:2019iul,Askar:2016jwt,Rodriguez:2016kxx} and direct $N$-body simulations~\cite{Banerjee:2016ths,DiCarlo:2019pmf}.} In Sec.~\ref{sec:massspin} we use this model to predict the fraction of events populating the mass and spin gaps.  In Sec.~\ref{1gerrors} we use simple analytical approximations for the effective spin probability distribution functions (PDFs) in field binaries and cluster binaries to estimate measurement errors on the ``mixing fraction'' between field and cluster events using only 1g mergers, and in Sec.~\ref{sec:frac} we show that using the mass and spin gaps can yield better estimates of the mixing fraction. In Sec.~\ref{sec:conclusions} we summarize our results and discuss directions for future research. The derivation of the spin PDFs is presented in Appendix~\ref{app:chieff}.
Throughout the paper, we use cosmological parameters from Ref.~\cite{Ade:2015xua}.
 
\section{Hierarchical mergers with a semianalytical cluster model}
\label{sec:clusters}

In this section we use a semianalytical cluster model which is not meant to replace $N$-body simulations, but serves our main purpose: relating the bulk properties of clusters to the characteristics of binary mergers which can be observed in GWs. 

\subsection{Binary formation and mergers in clusters}

Massive clusters are hotbeds for multiple-generation BBH mergers, but a good understanding of their evolution is elusive because the large number of particles comprising these systems makes numerical simulations extremely challenging. We evolve binary BHs in clusters following Refs.~\cite{Antonini:2016gqe,Choksi:2018jnq}. Reference~\cite{Antonini:2016gqe} used a semianalytical approach to predict rates and properties of inspiraling BH binaries forming in nuclear star clusters (NSCs), while Ref.~\cite{Choksi:2018jnq} combined a cosmological model of globular cluster (GC)
formation with analytical prescriptions from Ref.~\cite{Antonini:2016gqe} to study the properties of dense clusters that form merging BH binaries.

We calibrate the half-mass radius $\rh$ for GCs to fits of late-type galaxies \cite{Antonini:2016gqe,2016MNRAS.457.2122G}:  %
\be
\rh = \begin{cases}
\displaystyle
3\,{\rm pc} \;\;\qquad &{\rm if} \quad  \Mcl\leq M_{\rm NSC},
\\ 
\displaystyle
2.14 \left(\f{\Mcl}{10^6\Msun}\right)^{0.321}\!\!\!\!{\rm pc}  \;\;\qquad &{\rm if} \quad  \Mcl> M_{\rm NSC},
\end{cases} 
\label{eq:rh}
\ee
where we set $M_{\rm NSC}=2.87\times 10^6\,\Msun$ (slightly lower than the value $M_{\rm NSC}=5\times 10^6\,\Msun$ used in Ref.~\cite{Antonini:2016gqe}) to ensure continuity between the GC and the NSC regime.

The escape velocity from the cluster is~\cite{Antonini:2016gqe}
\be\label{eq:vesc}
v_{\rm esc}\simeq 0.1 \sqrt{\f{M_{\rm cl}}{M_\odot}
	\f{\rm pc}{r_{\rm h}}} \, \kms,
\ee
and the velocity dispersion  \new{is given by
$\sigstar = v_{\rm esc}/(2\sqrt{3})$, as predicted by Plummer's model}.
The number density of stars at the center of the cluster is set to \cite{Antonini:2016gqe}
\be
\nstar =4\times 10^6\left(\f{\sigstar}{100\,\kms} \right)^2\pcc
\ee
to match observations~\cite{Merritt:2009mr,Harris:1996kt}.

\subsubsection{Mass segregation}
\new{
Clusters containing a subpopulation of  BHs with average mass $\mbh$
will segregate to the cluster core on a timescale~\cite{1987gady.book.....B}}
\be\label{eq:tms}
\tms=\f{\mstar}{\mbh} t_r(\rh)\,,
\ee
where $\mstar$ is the mass of a typical star in the cluster and $t_r(\rh)$ is the relaxation time at the half-mass radius %
\be\label{eq:th}
t_r(\rh) = 4.2\times 10^9 \left(\f{15}{ \ln \Lambda} \right) 
\left(\f{\rh}{4\rm\ pc}\right)^{3/2}\left(\f{\Mcl}{ 10^7 \ M_\odot}\right)^{1/2} \rm \ yr\,.
\ee
We set the Coulomb logarithm parameter to %
 $\Lambda\simeq 0.4 N_\star$~\cite{1987gady.book.....B},
where $N_\star\simeq \Mcl/\mstar$ is the number of stars in the cluster.

As they fall into the core, BHs lose energy to stars, which become more energetic and migrate outwards. Over time the BHs become confined to an ever smaller core, where fewer stars are available to carry out the energy. Eventually  BHs decouple from the rest of the cluster population.  Assuming that the fraction of the total cluster mass contained in BHs is~\cite{Choksi:2018jnq}
\be
f_{\rm BH}=\f{\Mbh}{\Mcl}=0.05
\ee
and that BHs are confined in the ``BH half-mass radius''
\be
r_{{\rm BH}}\equiv\f{\Mbh}{\Mcl} \rh=f_{\rm BH} \rh\,,
\ee
one can find the number density of BHs as~\cite{Lee:1994nq}:
\bea
\nbh = \nstar\f{\Mbh}{\Mcl}\f{\rh^3}{r^3_{{\rm BH}}}
=  f_{\rm BH}^{-2}\, \nstar\,,
\eea
where $\nstar$ is the number density of stars in the core.

The velocity dispersion of BHs in this dynamically decoupled core is related to the stellar dispersion through the temperature ratio
\be\label{eq:temp_ratio}
\xi=\f{\mbh \sigbh^2}{\mstar\sigstar^2} = 5\,,
\ee
where for  the latest equality we follow Refs.~\cite{Choksi:2018jnq,AtakanGurkan:2003hm,Morscher:2014doa}.

\subsubsection{Formation of BH binaries}\label{BBH-formation}
In the dense environment of the cluster core BHs can efficiently form binaries, which will then harden and eventually merge through the following processes.

\begin{itemize}
\item[1)] {\bf Three-body interactions.} If the density is high enough, a close encounter between three single BHs can lead to the formation of a BH binary, with the third BH carrying away the energy needed to bind the pair. The timescale to form a binary via three-body interactions is~\cite{Lee:1994nq}
  \bea\label{eq:t3bb}
\tthreebb &=&  6.45\times 10^{9} \left(\f{\nbh}{ 10^6\,
	\pcc}\right)^{-2} \left({ \f{\sigbh}{10\,\kms}}\right)^{9}\nn \\
&&\times
\left( \f{m_1}{10\Msun} \right)^{-5}
\rm \ yr \,,
\eea
where $m_1$ is the mass of the heaviest BH in the triple system.  Three-body  binary formation is highly efficient because of the strong dependence on the velocity dispersion, which is much smaller for BHs compared to stars: cf. Eq.~(\ref{eq:temp_ratio}).

\item[2)]{\bf Binary-single interactions.} \new{Clusters also have a population of stellar binaries}, which tend to sink towards their cores because they are heavier than single stars. %
Once inside the core, these binaries undergo binary-single interactions with BHs. Most such encounters end up in exchanges between the BH and the lighter of the two stars in the binary. If the cluster core contains enough hard stellar binaries, a BH of mass $m_{\rm BH}$ can form a binary \new{with a star} via exchange interactions on a timescale~\cite{Miller:2008yw}
\bea\label{eq:t12}
\tonetwo &=& 5\times 10^9 \left( \f{f_{\rm b}}{0.1} \right)^{-1}\left(\f{\nstar}{ 10^4\,\pcc} \right)^{-1}\f{\sigbh}{10\,\kms} \nonumber \\
&& \times \left(\f{2\mstar + m_{\rm BH}}{ 20\Msun} \right)^{-1} \left(\f{\ahard}{
	1\rm\ AU}\right)^{-1}\rm \ yr \,,
\eea
where $f_{\rm b}=0.1$ is the binary fraction in the core~\cite{Ivanova:2005mi} and $\ahard$ is the typical separation of a hard stellar binary. The latter is estimated as~\cite{Heggie:1975tg,1989ApJ...343..725Q}:
\be\label{eq:ahard}
\ahard = 1.5 \left(\f{r_{\mathrm{h}}}{3\,\mathrm{pc}}\right)\left(\f{\mstar/\Mcl}{10^{-5}}\right)\,\mathrm{AU},
\ee
which corresponds to the maximum separation of a hard stellar binary in the core. %
\new{This BH--star system might form a BH--BH binary following another exchange
interaction with a single BH on timescales smaller than $\tonetwo$.} Comparing Eq.~(\ref{eq:t3bb}) and Eq.~(\ref{eq:t12}) shows that three-body binary formation is likely to dominate the  dynamical formation of BH binaries, because the binary fraction $f_{\rm b}$ is rather small. %

\item[3)]{\bf GW captures.} BHs can also form binaries through single-single GW capture. In this case, two single BHs become bound after a close encounter if sufficient energy is dissipated via GWs.  Such BH binaries are predicted to be very eccentric~\cite{Samsing:2019dtb} and, consequently, merge almost instantly \cite{Peters:1963ux}. The rate of single-single GW capture mergers is comparable to that of binary-single interactions only when the binary fraction is at the percent level~\cite{Samsing:2019dtb}.  For $f_b=0.1$ as assumed here, GW captures can be safely neglected. %
\end{itemize}

To summarize, we define the BBH formation timescale to be $\min(\tonetwo, \tthreebb)$.

\subsubsection{Hardening and Merger}
After a ``hard'' binary (i.e., a binary with binding energy greater than the kinetic energy of cluster particles) %
 is formed, it typically undergoes a series of strong encounters with stars in the core. Because of ``Heggie's law''~\cite{Heggie:1975tg}, these repeated encounters tend to make hard binaries harder and soft binaries softer. In a cluster with  BH \new{mass density} $\rhobh$, a binary will harden at a rate~\cite{Quinlan:1996vp} %
\be\label{eq:adotdyn}
\dot{a}_{\rm dyn} = -20 \f{G\rhobh}{\sigbh^2}a^2\,.
\ee
If, after an interaction with another BH, the semimajor axis $a$ of the binary decreases to $a_{\rm fin}$, the binary
will recoil with velocity $\propto(a_{\rm fin})^{-1/2}$. This happens because the extra binding energy is converted to kinetic energy, most of which is carried away by the interloper, while some of it gets transferred to the binary system.

So, while binaries become harder with every encounter, these binaries also receive larger and larger recoils, and may eventually be kicked out of the cluster. By equating the recoil speed to the escape velocity of the cluster $v_{\mathrm{esc}}$, one can estimate the binary separation at which the binary could be ejected:
\be\label{eq:aej}
\aej = 3.9\,\eta\left(\f{v_{\mathrm{esc}}}{30\,\kms} \right)^{-2} \f{m_3}{20 \Msun}  \f{m_3}{m_1+m_2+m_3}\,\mathrm{AU}\,,
\ee
where $m_1$ and $m_2$ are the BBH component masses, $\eta\equiv m_1 m_2/(m_1+m_2)^2$ is the symmetric mass ratio of the BBH, and $m_3=\mbh$ is the mass of the BH interloper.

A binary can avoid ejection if gravitational radiation takes over and drives it to merger {\em before} another dynamical interaction kicks it out of the cluster. The binary separation decays due to gravitational radiation as
\be\label{eq:adotGW}
\dot{a}_{\rm GW} =
-\f{64}{ 5}\f{G^3}{c^5}\f{m_1m_2 (m_1+m_2)}{a^3}f(e)\,,
\ee
where $e$ is the eccentricity and
\be
f(e)=\left(1+\f{73}{24}e^2+\f{37}{96}e^4\right) (1-e^2)^{-7/2}\,.
\ee
We adopt the median value $e = 1/\sqrt{2}$ expected for a thermal distribution $p(e) = 2e$. GWs start dominating the dynamics at the separation $a_{\rm GW}$ where dynamical hardening [Eq.~(\ref{eq:adotdyn})] balances GW emission [Eq.~(\ref{eq:adotGW})]. Setting $\dot{a}_{\rm dyn}=\dot{a}_{\rm GW}$ yields %
\begin{align}
\label{eq:agw}
\agw ={}& 0.05 \left(\f{m_1+m_2}{
	20\,\Msun}\right)^{3/5} \nonumber \\
&\! \times\left[ \eta \f{\sigbh}{30\, \kms} \f{ 10^6 
	M_{\odot}\!{\rm\ pc}^{-3}}{ \rhobh} f(e)\right]^{1/5}
\rm AU  \,.
\end{align}

For $a<a_{\rm GW}$, GWs dominate the energy loss from the binary. If $\agw < \aej$, \new{dynamical} encounters eject the BBH from the cluster before GWs can drive the BHs to coalescence. In this case, the ejected BBH (with a separation $\aej$) can continue to harden ex situ via GW emission. On the other hand, if $\agw > \aej$, GW emission will cause the BBHs to coalesce in situ before ejection.

Assuming that each interaction extracts 20\% of the binary's binding energy~\cite{Quinlan:1996vp}, the time to harden to a separation $a_{\mathrm{crit}} = \mathrm{max}(a_{\mathrm{GW}}, a_{\mathrm{ej}})$, from an initial separation $a \gg a_{\rm crit}$ is~\cite{Miller:2001ez}
\bea
\thard &=&  \f{3.16 \times 10^8}{q_3}\left(\f{\sigbh}{30\,\kms}\right)  \left(\f{\acrit}{0.05 \rm AU}\right)^{-1}
\nonumber \\ &&\times \left(\f{m_1+m_2}{20 \Msun} \right)^{-1}   \left( \f{\nbh}{10^6\,\pcc}\right)^{-1} \, \mathrm{yr}.
\label{eq:tcrit}
\eea
The binary will continue to interact with other  cluster members until it reaches a
semimajor axis $a_{\rm crit}$.
After reaching $\acrit$, the binary's hardening is dominated by GW emission, which drives the system to coalescence on a timescale \cite{Peters:1963ux} %
\bea
\label{eq:tgw}
\tgw &=& 0.56 \times 10^{8} \left[\f{m_1m_2 (m_1+m_2)}{2\times10^3 \Msun^3}\right]^{-1} \notag
\\
&&\times\left(\f{\acrit}{0.1 \mathrm{AU}}\right)^4 \! \left(\f{1-e^2}{0.5} \right)^{7/2}\mathrm{yr}\,.
\eea

\begin{figure}[t]
  \includegraphics[width=\columnwidth]{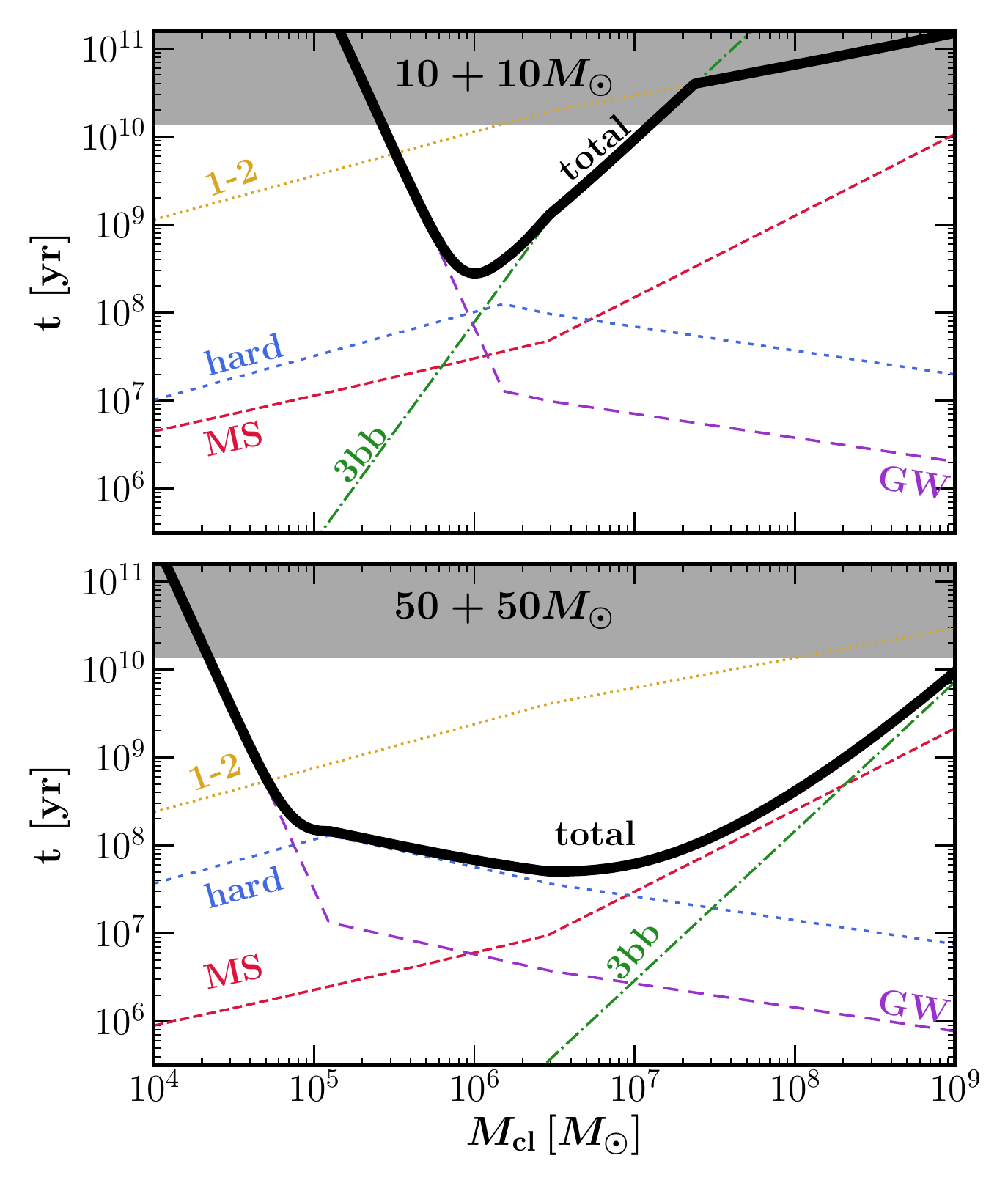}
  \caption{Timescales involved in the merger of $(10+10)\Msun$ (top) and $(50+50)\Msun$ (bottom) BBHs. The timescales related to three-body interactions, binary-single interactions, mass segregation, GW inspiral, and critical hardening are indicated in green, yellow, red, purple, and blue, respectively. The thick black line marks the sum $\ttot$ of Eq.~(\ref{ttot}). The gray shaded region marks time delays larger than the age of the Universe.}
\label{fig:ttotal2}
\end{figure}

\subsubsection{Timescale comparison}

The total delay time between the formation of the cluster and a BBH merger is the sum of the timescales for mass segregation [Eq.~(\ref{eq:tms})], BBH formation 
[$\min(\tonetwo, \tthreebb)$], hardening [Eq.~(\ref{eq:tcrit})] and GW-induced merger [Eq.~(\ref{eq:tgw})]:
\be
\ttot = \tms+\min(\tonetwo, \tthreebb)+\thard+\tgw\,.
\label{ttot}
\ee
\new{Our estimates neglect the lifespans of massive stars that lead to BH formation, which is of order $\mathcal{O}(1) {\rm Myr}$.}
Figure~\ref{fig:ttotal2} shows how the different terms in this sum depend on $\Mcl$:

\begin{itemize}
\item[1)] \textbf{Mass segregation.} BHs sink into the core on a timescale $\tms\propto {(\rh^3 \Mcl)^{1/2}}$ [Eq.~(\ref{eq:tms})]. In our model GCs have a fixed $\rh$, so $\tms\propto \Mcl^{1/2}$; for NSCs the mass segregation timescale is approximately $\tms\propto\Mcl$ [Eqs.~(\ref{eq:rh}) and (\ref{eq:th})]. 

\item[2)] \textbf{BBH formation.} BH binaries form predominantly through three-body interactions, which are very sensitive to the velocity dispersion [Eq.~(\ref{eq:t3bb})]. Heavier clusters have a larger velocity dispersion which makes three-body interactions inefficient, therefore BBH formation timescales increase very steeply with cluster size ($\tthreebb\propto\Mcl^{5/2}$). As a result binary formation through three-body interactions is slower in most NSCs compared to GCs, where BBHs could also form through binary-single interactions on a timescale $\tonetwo\propto\Mcl^{1/6}$ [Eq.~(\ref{eq:t12})]. %

\item[3)] \textbf{Hardening.} For ex situ mergers, binaries are ejected more rapidly for smaller clusters, and therefore they spend less time hardening ($\thard\propto\Mcl^{1/2}$). On the other hand, the hardening timescale for clusters that retain their binaries decreases with cluster mass ($\thard\propto\Mcl^{-2/5}$): larger clusters have higher densities and a larger influx of BHs to the center, which makes dynamical hardening more efficient.

\item[4)] \textbf{GWs.} Small clusters have low escape speeds and BBHs get ejected at large orbital separations ($\aej\propto\Mcl^{-1}$) due to dynamical interactions. Therefore, the gravitational radiation reaction timescale $\tgw\propto \acrit^4\propto\Mcl^{-4}$ increases sharply for small clusters. The situation is different for heavier clusters, which retain and dynamically harden BBHs until GW emission takes over: in this case $\agw\propto\Mcl^{-1/10}$, and thus $\tgw\propto\Mcl^{-2/5}$.
\end{itemize}

The dominant term depends on both the mass of the cluster and the mass of the binary (Fig.~\ref{fig:ttotal2}). For lighter BBHs of $\sim(10+10)\Msun$, only the GW timescale matters for clusters with mass $\lesssim 10^6 M_\odot$, while the three-body interaction timescale is dominant for large clusters. For larger BBHs of $(50+50)\Msun$,  hardening time and mass segregation timescales also play an important role, while the GW radiation-reaction timescale and the three-body timescale become important only  for $\Mcl \lesssim 10^5 M_\odot$ and  $\Mcl \gtrsim 10^8 M_\odot$, respectively. Figure~\ref{fig:ttotal2} also confirms our earlier claim that binary-single interactions are not an efficient channel for BBH formation. %
They only become important for light binaries in very massive clusters, but at that point $\tonetwo$ becomes comparable to the age of the Universe.

Figure~\ref{fig:ttotal2} also shows the total delay time %
for equal-mass binaries as a function of $\Mcl$. %
For a binary with fixed component masses $m_1=m_2$, the delay time decreases with $\Mcl$ for clusters of mass $\Mcl\lesssim 10^6\Msun$, where the gravitational radiation or hardening timescales dominate. For $\Mcl\gtrsim 10^6\Msun$, other processes dominate and the delay time increases.
The minimum time delay $\ttot^{\rm min}$ over all cluster masses is, in general, a function of $m_1$ and $m_2$:
\bea
\ttot^{\rm min}(m_1,m_2) &=& \min [ \ttot(m_1,m_2,\Mcl) ]\,,
\eea
where we take $\Mcl \in [10^5 \Msun,10^9\Msun]$.
For a given binary of masses $(m_1,\,m_2)$, as long as $t_0>\ttot^{\rm min}$, there are two values of $\Mcl$ -- say  $\Mcl^{\rm min}$ and $\Mcl^{\rm max}$ -- such that 
\bea\label{eq:Mclminmax}
\ttot(m_1,m_2,\Mcl^{\rm min})=t_0\,,\nn\\
\ttot(m_1,m_2,\Mcl^{\rm max})=t_0\,.
\eea
These represent bounds on the range of cluster masses that can produce merging BHs with masses $m_1$ and $m_2$ within time $t_0$: BBH mergers are possible, \new{on average,} when $\Mcl^{\rm min}\le\Mcl\le\Mcl^{\rm max}$. This point will be important later (cf. Sec.~\ref{sec:mcl}).

\begin{figure}[t]
  \includegraphics[width=\columnwidth]{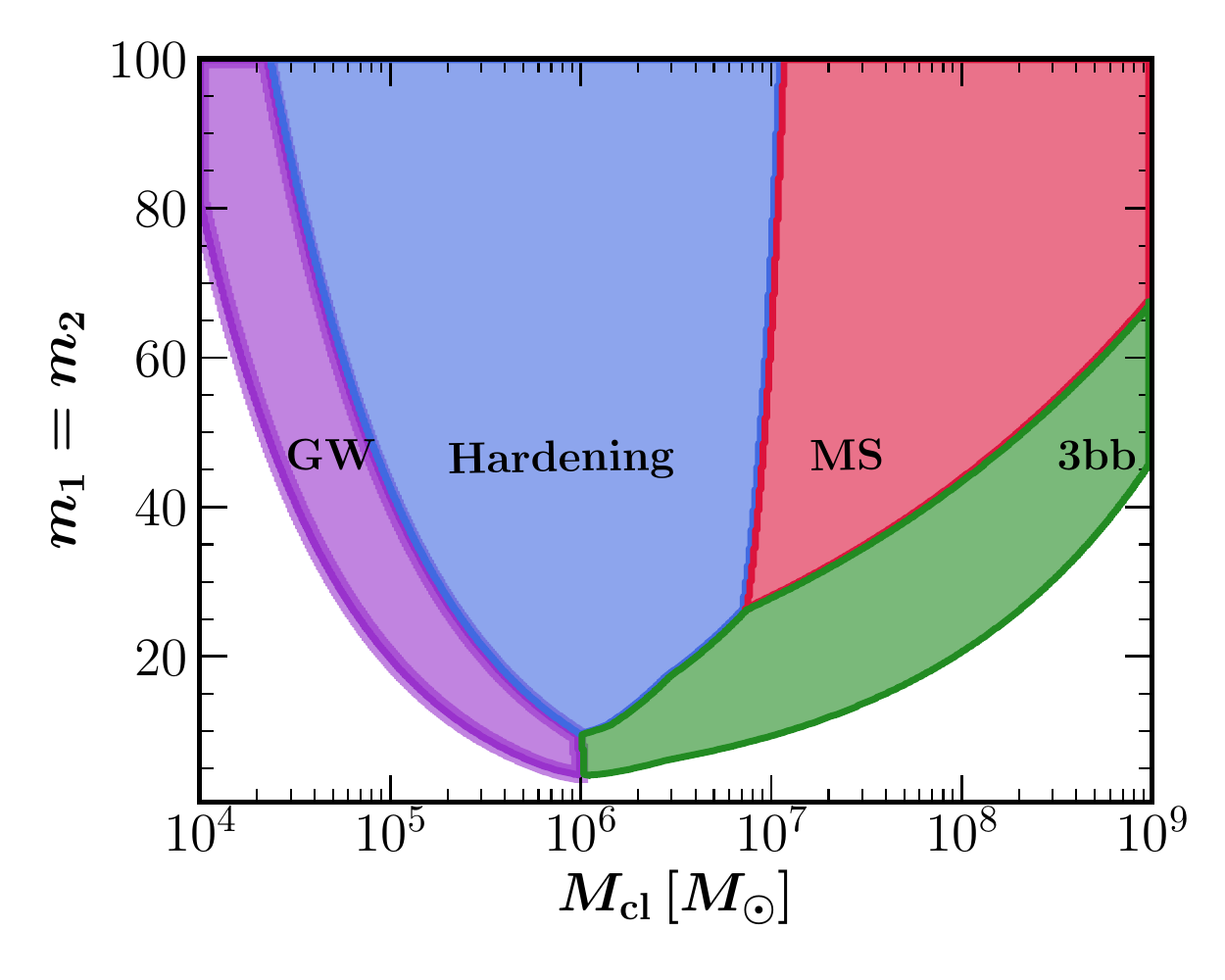}
  \caption{Dominant timescales in the $(\Mcl, m_1)$ plane for an equal-mass binary $(m_1=m_2)$. Regions where three-body interactions, mass segregation, hardening, and GW inspiral dominate are indicated in green, red, blue, and purple, respectively.} 
\label{fig:ttotal3}
\end{figure}

Figure~\ref{fig:ttotal3} illustrates the dominant timescale as a function of cluster mass and BBH mass for an equal-mass binary ($m_1=m_2$).
Most BBHs have small masses (cf. Sec.~\ref{sec:1g}), where $\ttot$ is dominated by $\tgw$ and $\tthreebb$~\cite{Choksi:2018jnq}. Both of these timescales decrease  sharply with the BH mass  ($\tgw \propto m_1^{-7}$ and $\tthreebb \propto m_1^{-5}$), so mass segregation and hardening (which decay as  $m_1^{-1}$ and $m_1^{-1.6}$, respectively) can become the dominant timescales only for  BBHs with larger masses. The hardening time dominates for $m_1 \gtrsim 10\Msun$, while the mass segregation timescale dominates for $m_1 \gtrsim 25\Msun$ in some NSCs with mass $\Mcl\gtrsim 10^{7} \Msun$. 

\begin{figure}[t]
  \includegraphics[width=\columnwidth]{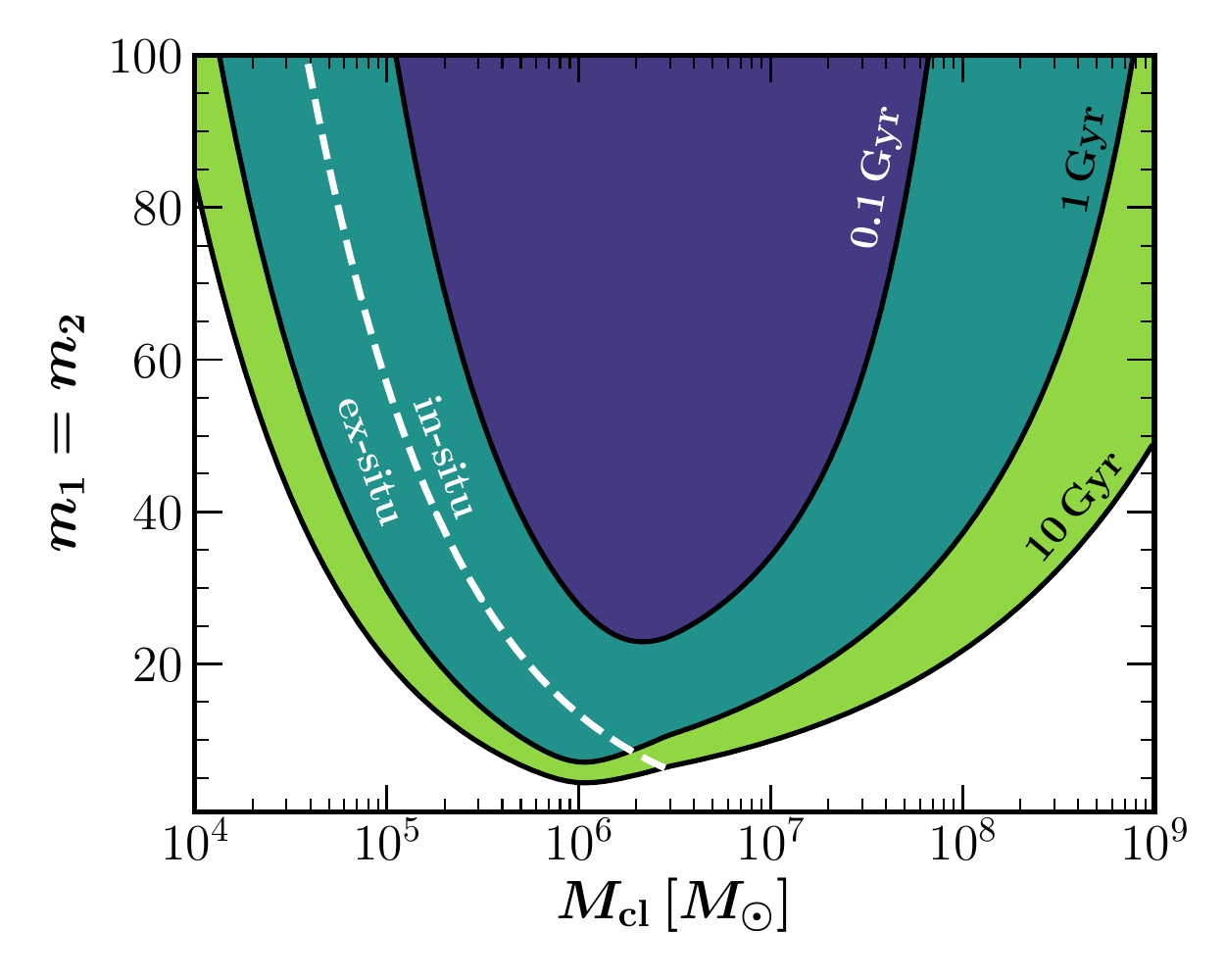}
  \caption{Contour plot of the total delay time for the merger of an  equal-mass BBH system $(m_1=m_2)$ as a function of the cluster mass. The white dashed line is the boundary between the regions where most mergers happen ex situ (left) and in situ (right).}
\label{fig:ttotal1}
\end{figure}

\begin{figure*}[t]
  \includegraphics[width=0.475\textwidth]{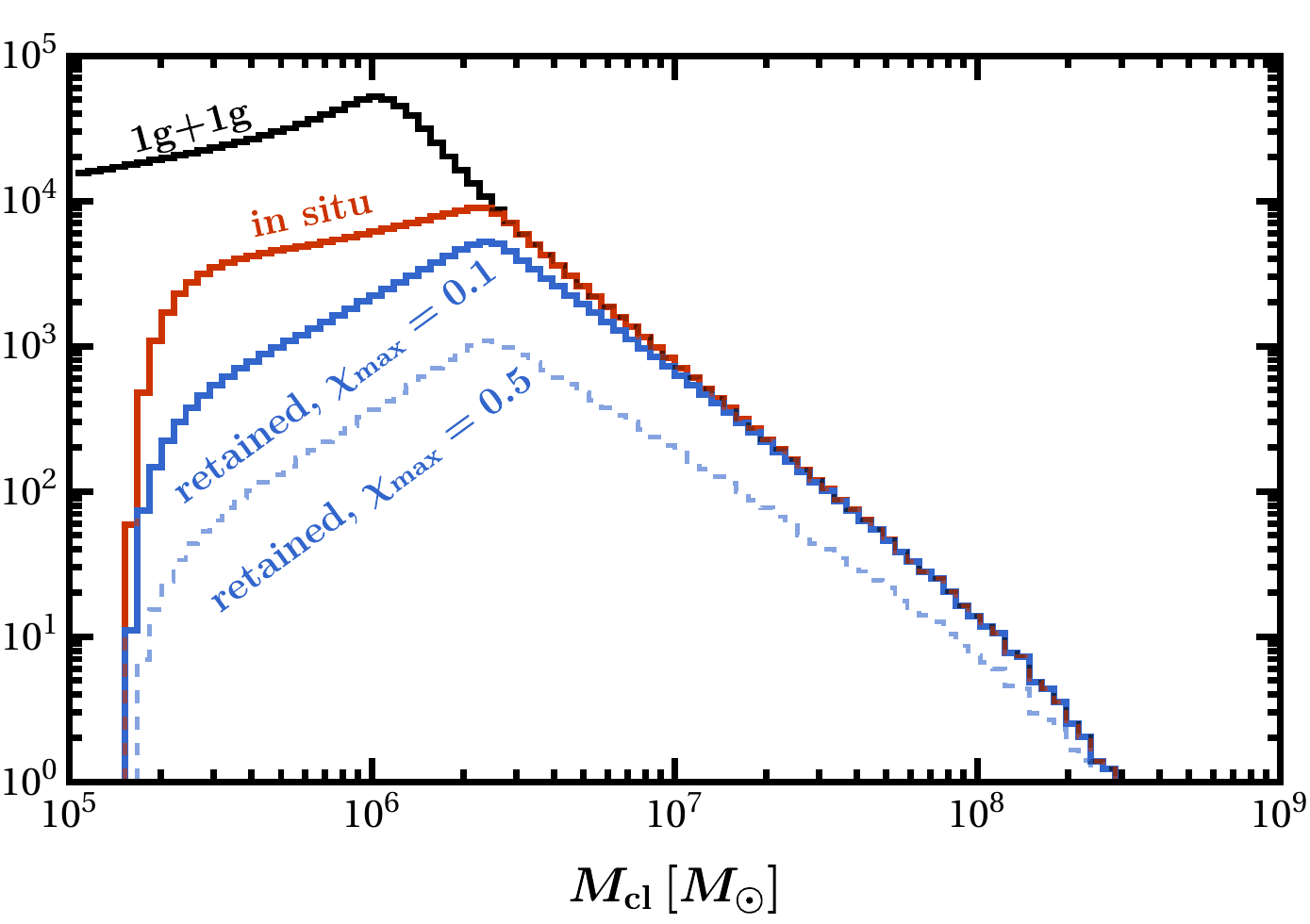}
  \includegraphics[width=0.475\textwidth]{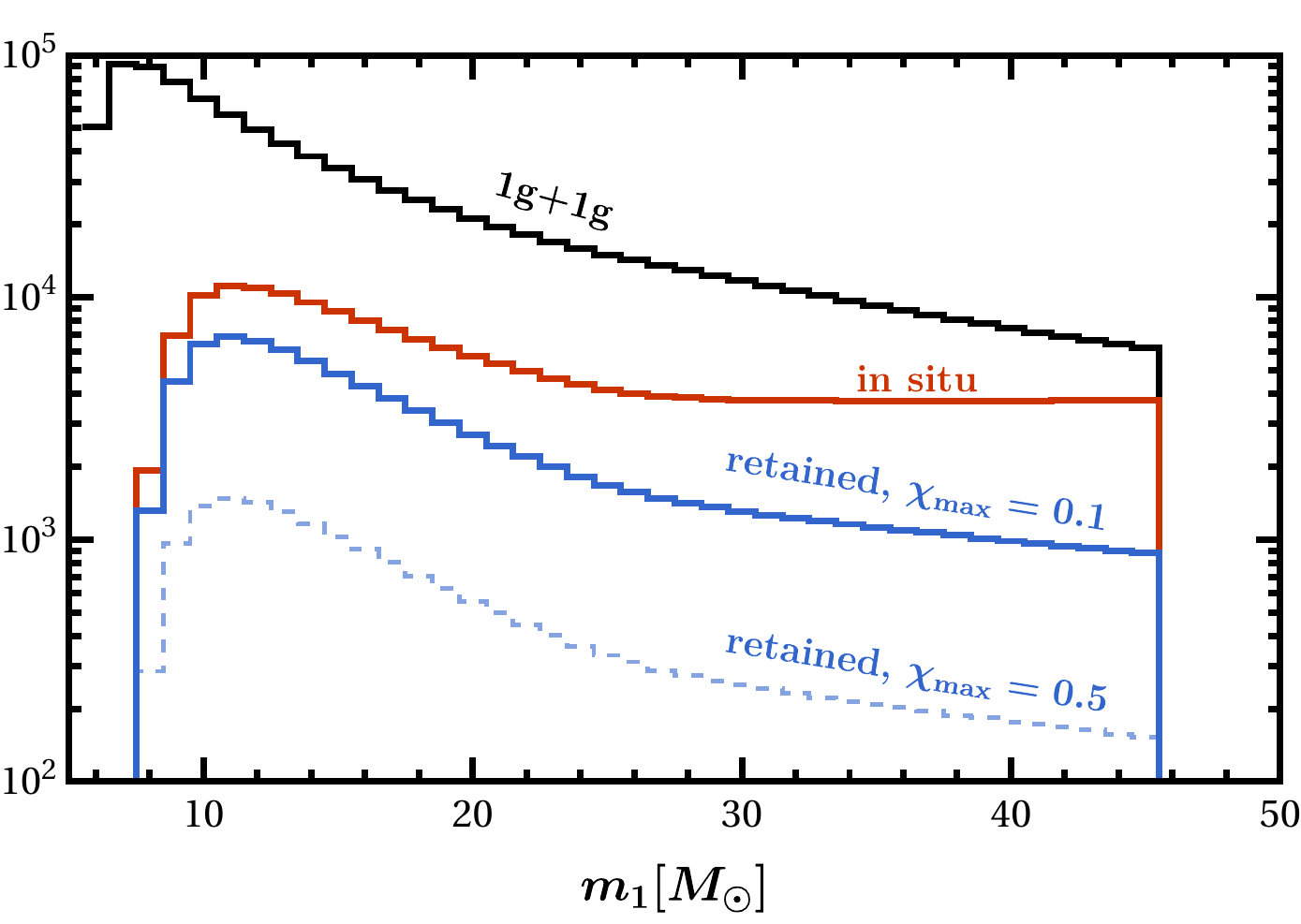}
  \caption{Distribution of host cluster masses (left) and primary BBH masses (right) for the 1g+1g populations. Black curves show the full sample of 1g mergers (higher generations  are excluded from this plot). Red curves show the fraction of binaries that survive dynamical kicks and are able to merge inside the cluster. Blue curves show systems that further survive GW kicks and remain available to assemble the second generation of BH mergers. We assume $\chimax=0.1$ (solid)  and $\chimax=0.5$ (dashed).} 
\label{fig:Mcl}
\end{figure*}

Figure~\ref{fig:ttotal1} shows how the time $\ttot$ varies in the $(\Mcl,\,m_1)$ plane, assuming again an equal-mass binary ($m_1=m_2$). In particular, the white dashed line separates the region where mergers occur in situ and ex situ: small BHs and BBHs hosted in small clusters are ejected because of large recoils and insufficient escape speed, respectively. After they are ejected, these BHs then take a long time to merge under gravitational radiation reaction: as discussed earlier, this phase dominates their entire evolution. More quantitatively, BHs with $m_1\lesssim 7 \Msun$ take $\gtrsim 10^9$ yr to merge. Our model does not predict mergers of BHs with $m_1\lesssim 4 \Msun$, which are always ejected before a binary is formed. We also find that below a minimum primary BBH component mass for in situ mergers
\be\label{eq:mmininsitu}
m_{\rm min,is} \simeq 6 \Msun
\ee
all mergers happen outside the cluster.

\subsection{First-generation mergers}
\label{sec:1g}

Ideally, one should generate the population of merging binaries by convolving a cluster formation model with the delay times discussed in the previous section. %
There are large uncertainties in this process \cite{2019MNRAS.482.4528E,Fragione:2018vty,2018RSPSA.47470616F}, so we choose instead to start by considering the observed population of first-generation BBHs. %

We distribute the primary mass of 1g+1g mergers according to \cite{LIGOScientific:2018jsj}
\be
p(m_1)\propto m_1^\alpha, \quad\quad m_1\in[\mmin,\mmax]\,,
\label{eq:pm1}
\ee
while the secondary mass is drawn from
\be
p(m_2|m_1) \propto m_2^\beta , \quad\quad m_2\in[\mmin,m_1]\,.
\label{eq:pm2}
\ee
We fix $\alpha=-1.6$  and  $\beta=6.7$, as estimated from GW observations~\cite{LIGOScientific:2018jsj}, and $\mmin =5\Msun$. The parameter $\mmax$ marks the onset of the mass gap and it is set to $45\Msun$~\cite{Farmer:2019jed} unless specified otherwise.
The spin magnitudes of the component BHs are drawn from a uniform distribution in the range $[0,\,\chimax]$.

We sample redshifts from the Madau star-formation-rate fit~\cite{Madau:2014bja}:
\be\label{eq:SFR}
{\rm sfr}(z) =  \frac{0.015(1+z)^{2.7}}{1 + [(1+z)/2.9]^{5.6}}\,. 
\ee
Lighter BHs have longer total delay times $\ttot$. Therefore, at any given redshift there is a lower bound on the masses of BBHs that could form and merge within a given time. Assuming that clusters could not have formed earlier than $t_{\rm max}=13.4\times 10^9$~years ago ($z_{\rm max}=11.34$), we discard all binaries that could not have merged at the sampled redshift. In other words, we only keep binaries that satisfy the constraint
\be
\tlb(z)+\ttot^{\rm min}(m_1, m_2)<t_{\rm max}\,,
\ee
where $\tlb$ is the cosmological lookback time~\cite{Hogg:1999ad}.  This procedure removes some low-mass and/or high-$z$ binaries, slightly modifying our merging population relative to the sampled distribution.

Starting from this 1g+1g population, we can now use the semianalytical scheme outlined above to obtain the distribution of $2\text{g}+1\text{g}$ and $2\text{g}+2\text{g}$ binaries.

\subsubsection{Clusters that merge black holes}
\label{sec:mcl}

For each binary we sample the cluster mass from a distribution of the form~\cite{Harris:1994eu}
\be\label{eq:Mcl_distribution}
p(\Mcl)\propto \Mcl^{-2}
\ee
in a range of cluster masses $\Mcl \in [\Mcl^{\rm min},\Mcl^{\rm max}]$ that could support BBH mergers at redshift $z$. Here $\Mcl^{\rm min}$ and $\Mcl^{\rm max}$ are calculated from Eq.~(\ref{eq:Mclminmax}), with $t_0=t_{\rm max}-\tlb(z)>\ttot$.

The black line in the left panel of Fig.~\ref{fig:Mcl} shows the resulting distribution of cluster masses.  Smaller clusters are more abundant, but relatively inefficient at bringing binaries to merger: BBHs get ejected from the cluster with large orbital separations, and therefore have long GW-driven inspiral timescales $\tgw$. %
Most of the merging binaries come from GCs with mass $\sim 10^6\Msun$: these clusters lie in the ``sweet spot'' where delay times are smallest, especially for lighter BBHs, which form the bulk of the population (cf. Figs.~\ref{fig:ttotal2}-\ref{fig:ttotal1}). This behavior is consistent with Refs.~\cite{Choksi:2018jnq,Antonini:2019ulv}. %
In our model, smaller BHs of component masses $\sim 5 M_\odot$  can only form in clusters with $\Mcl\simeq10^{5.8} - 10^{6.2} \Msun$: these are the only systems that can efficiently lead light BBHs to merger. On the other hand, heavier BBHs have smaller delay times, and they can merge more easily within a wider range of cluster masses. Because of the shape of the probability distribution function $p(\Mcl)$, most of these massive BBHs come from clusters at the lower end of the mass spectrum.

\subsection{Hierarchical mergers}
\label{hiemer}
Our goal is to look for smoking guns that can be used to identify the 2g population.
We must first address a key question: how efficiently do clusters produce 2g mergers? 

\subsubsection{Retention in the cluster}
\label{sec:retention}
Given a sample of binaries with masses $(m_1,\,m_2)$ merging in a cluster of mass $\Mcl$, we assume that all binaries with $\agw<\aej$ merge inside the cluster. At merger, the remnant receives an additional kick $\vkick$ due to asymmetric GW emission. If $\vkick<\vesc$ the remnant is retained in the cluster, where it can merge again and form a 2g binary.  The properties of the merger remnant are computed using fits to numerical-relativity simulations for final mass~\cite{Barausse:2012qz}, spin~\cite{Hofmann:2016yih} and recoils~\cite{Campanelli:2007ew,Gonzalez:2006md,Lousto:2007db,Lousto:2012su,Lousto:2012gt} as implemented in Ref.~\cite{Gerosa:2016sys}, assuming isotropic spin orientations.

\begin{figure}[t]
  \includegraphics[width=0.49\textwidth]{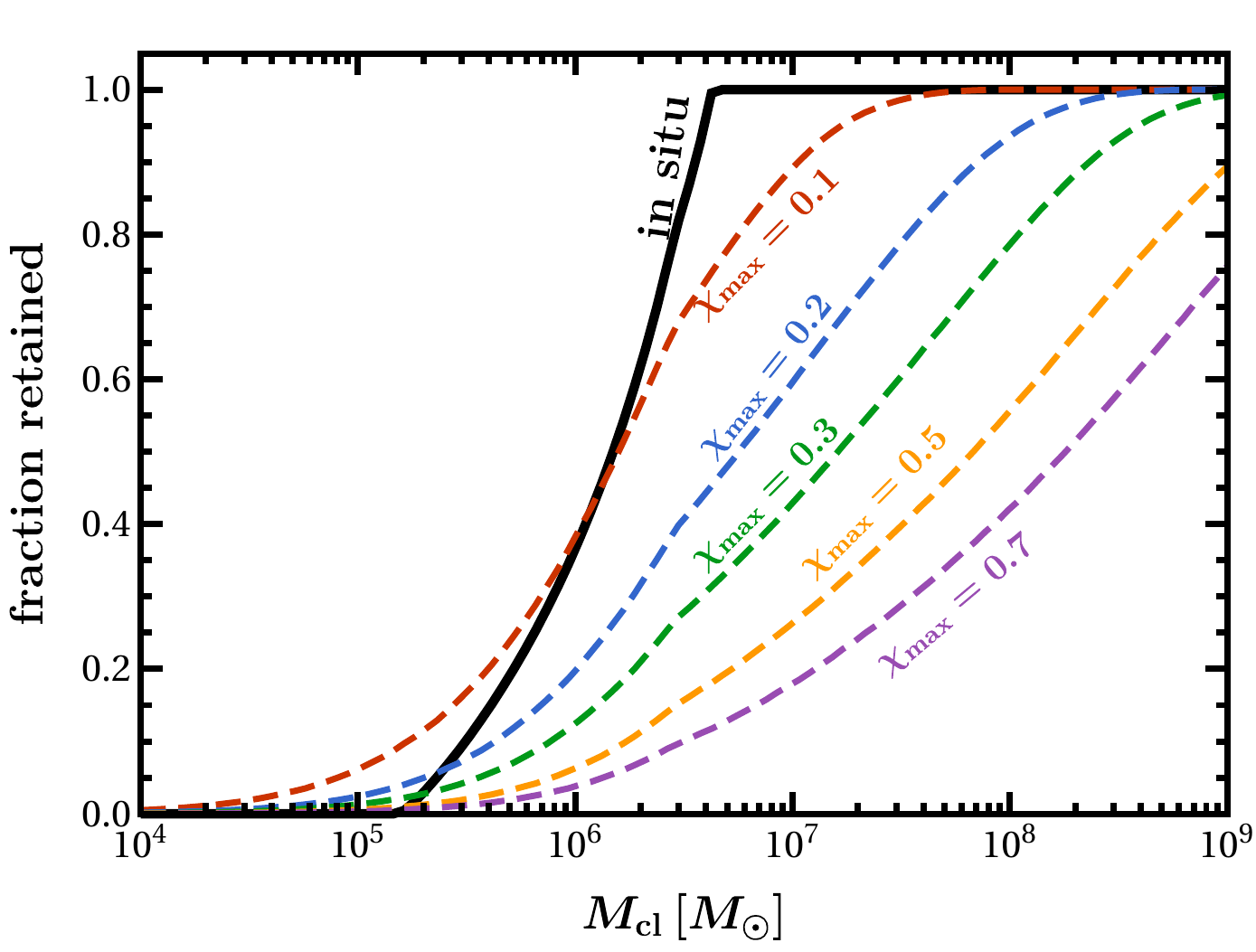}
  \caption{ Fraction of BBHs retained after kicks due to either (i) dynamical interactions before merger (solid line), or (ii) GW recoil at merger (dashed lines). The largest spin of 1g BHs $\chi_{\rm max}$ increases from top (red, $\chi_{\rm max}=0.1$) to bottom  (purple, $\chi_{\rm max}=0.7$). %
  } 
\label{fig:frac_ret}
\end{figure}

For illustrative purposes, in Fig.~\ref{fig:frac_ret} we focus only on the retention power of clusters ignoring the prescriptions of Sec.~\ref{sec:mcl}, as well as the fact that some clusters might not be able to drive small BHs to merger. We plot both the fraction of in situ mergers ($\aej<\agw$) and retained BHs ($\vkick<\vesc$) for different assumptions on the quantity $\chi_{\rm max}$  that marks the ``edge'' of the spin gap. in situ BH mergers are only possible for clusters with mass $\Mcl>1.6\times10^5 \Msun$, where this threshold is mainly set by $\mmax$. For $\Mcl>4\times10^6 \Msun$ -- a threshold now set by $\mmin$ -- all mergers are found in situ. %
Post-merger kicks increase when the merging BH spins are large, so the retention fraction decreases steeply with $\chimax$.  
For larger values of $\chimax$, the population contains more BHs that receive kicks larger than the escape speed of all but the most massive clusters.

In the right panel of Figure~\ref{fig:Mcl} we convolve this retention power and the prescription of Sec.~\ref{sec:mcl} to illustrate the final distribution of BBH primary masses $m_1$.
The 1g+1g distribution of primary masses follows the injected power law $m_1^{-1.6}$. Most small-mass BHs get ejected due to kicks from dynamical interactions, resulting in an almost flat distribution for in situ mergers with $m_1\gtrsim 30 M_\odot$  In addition, in situ mergers do not occur for $m_1<m_{\rm min,is}\simeq 6\Msun$ (cf. Figs.~\ref{fig:ttotal1} and~\ref{fig:Mcl}).

The mass distribution of the retained binaries depends on $\chimax$. Larger $\chimax$ leads to large kicks: this reduces the number of retained remnants, but also affects the slope of the distribution. This can be understood as follows. Because $\beta\gg1$, most binaries have mass ratio $q$ close to unity.  For $\chi_{\rm max}=0$, symmetry in the merger process (e.g. \cite{Gerosa:2018qay}) implies $\vkick=0$. In this case, the $m_1$ distributions of in situ and retained mergers should be very similar, with the same plateau at $m_1\sim m_{\rm max}$. For larger values of $\chimax$, however, most of the small clusters do not retain post-merger remnants, which removes a significant fraction of the heaviest BBHs. These systems can form and merge easily in a wider range of cluster masses, which implies that a significant fraction of them comes from the lighter, more abundant environments. Therefore, large values of $\chimax$ lead to an $m_1$ distribution of retained BBHs which drops more sharply.%

The left panel of Fig.~\ref{fig:Mcl} shows the mass distribution of clusters that host BBHs. As expected, in situ and retained cluster events are more likely in heavier systems with a larger escape speed. In particular, clusters with $\Mcl\gtrsim 3\times10^6\Msun$ are able to produce in situ events. The cluster masses that can retain BHs following GW kicks and thus support 2g mergers are sensitive to the maximum spin $\chimax$ of 1g BHs. For $\chimax\gtrsim 0.5$, some BHs are ejected even from the most massive NSCs. %

\subsubsection{2g+1g or 2g+2g?}
\label{sec:2gvs1g}

If retained, a BBH merger remnant (a ``2g BH'') can merge with either another 2g BH (2g+2g merger) or with a 1g BH (2g+1g merger). %
Selective pairing of BH component masses implies that the retained remnants of 1g mergers are more likely to merge with a 2g BH (cf.~\cite{Gerosa:2019zmo}). However, because heavier BHs merge more quickly, 2g systems have a very small survival time in the cluster compared to 1g BHs. Although  2g BHs would tend to pair with other 2g BHs, their short merger time implies a lower merger probability for 2g+2g mergers. %
 
 In order to take into account some of these complications, we assign a merger probability based on (i) the number density of 1g BHs, and (ii) the number density of BHs retained within a given cluster at a given redshift. We divide the 1g population into bins of cluster mass and redshift, ${\mathfrak n}_{1 \rm g}(\Mcl,\,z)$. We do the same for the retained remnants and calculate the number of binaries in each bin, ${\mathfrak n}_{\rm rem}(\Mcl,\,z)$. The number of ways in which a 2g+1g binary can form is proportional to ${\mathfrak n}_{1 \rm g}\times {\mathfrak n}_{\rm rem}$, while for a 2g+2g binary it is proportional to $ {\mathfrak n}_{\rm rem}({\mathfrak n}_{\rm rem}-1)/2$. So the ratio between 2g+2g and 2g+1g mergers in a cluster of mass $\Mcl$ at redshift $z$ is $\propto ({\mathfrak n}_{\rm rem}-1)/(2 {\mathfrak n}_{1 \rm g})$. %

If a 2g BH of mass $m_1$ merges with another 2g BH, we  extract the companion mass $m_2\in[\mmin,\,m_1]$ from the same distribution $p(m_2|m_1)\propto m_2^\beta$ used in Sec.~\ref{sec:1g}. Its spin is estimated by binning and resampling the 2g remnant spin distribution. For the case of 2g+1g mergers, we extract $m_2$ from $p(m_2|m_1)\propto m_2^\beta$ but now restrict 
$m_2\in[\mmin, \min(m_1,\mmax)]$. The spin is extracted from the 1g distribution $p(\chi)=$ constant with $\chi\in[0,\chimax]$.

We calculate the time delay between a 1g+1g merger and the next merger involving its 2g remnant using the same $\ttot$ introduced in Eq.~(\ref{ttot}) above, but replacing the mass segregation timescale $\tms$ by $\tms\times (\vkick/\vesc)^3$ to take into account the time needed by the kicked remnant to sink back into the cluster core. This is obtained from Eq.~(\ref{eq:tms}) by assuming that the GW recoil displaces the remnant to $(\vkick/\vesc)^2\rh$~\cite{Antonini:2016gqe}. %

\begin{figure}
   \includegraphics[width=\columnwidth]{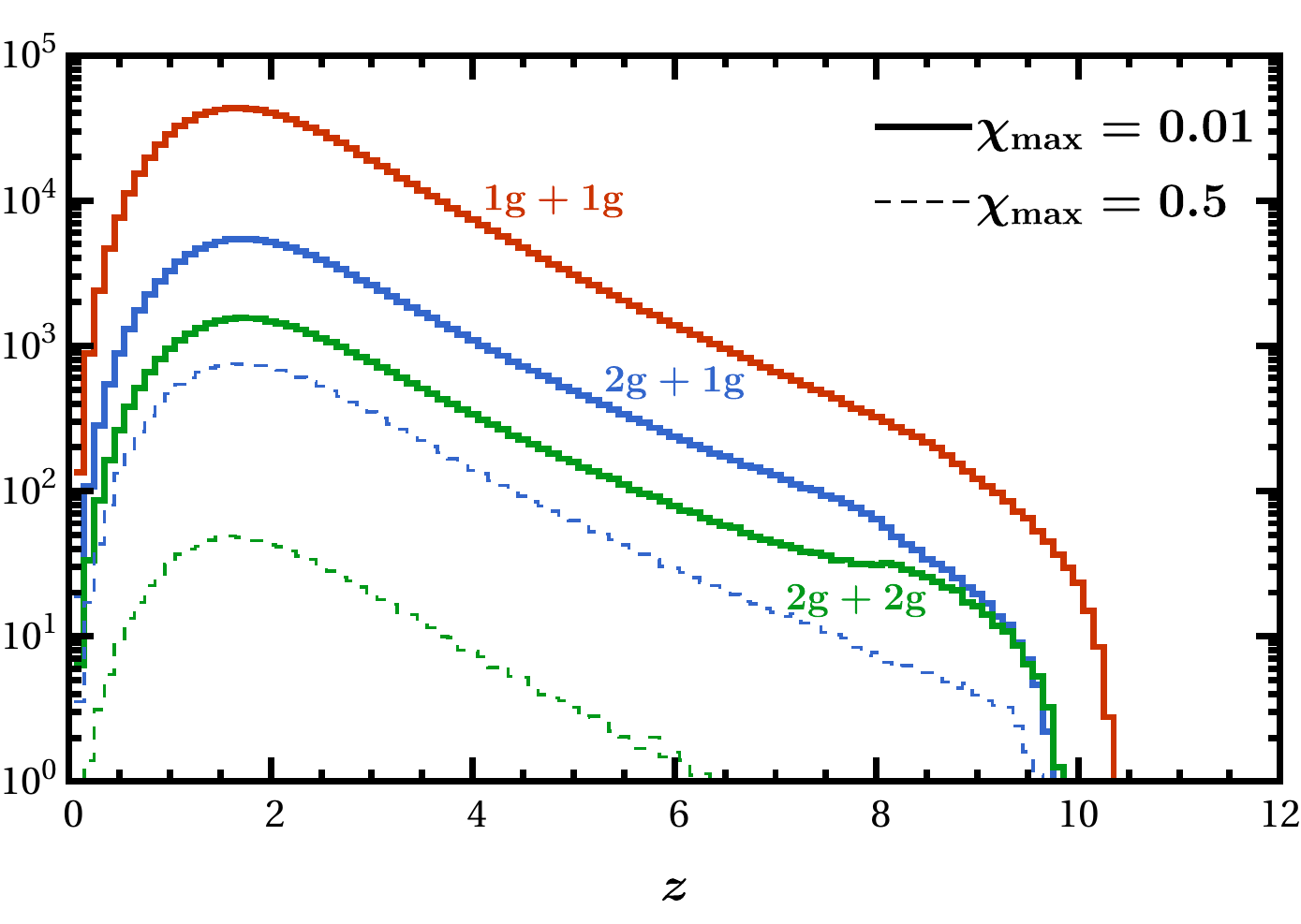}
  \caption{Distribution of merger redshifts $z$.  Red, green, and blue curves indicate the 1g+1g, 2g+1g, and 2g+2g populations, respectively. Solid (dashed) histograms are obtained with $\chi_{\rm max}=0.01$ ($\chi_{\rm max}=0.5$).} 
\label{fig:z}
\end{figure}

The redshift distribution of 1g+1g mergers was assumed to follow the Madau star formation rate of Eq.~(\ref{eq:SFR}). We find that 2g BBHs closely follow the same distribution, contrary to the expectation that repeated mergers may suffer further time delays~\cite{Gerosa:2017kvu}: cf. Fig.~\ref{fig:z}. This is because merger products are heavier compared to their progenitors, and thus merge  on very short timescales in our model.

We find that, overall, the number of 2g+1g events dominates over the 2g+2g populations with a relative fraction of about 4:1 for $\chimax=0.01$. The fraction of 2g+2g mergers decreases and tends to zero as $\chimax \to 1$.
 
 \new{
 \subsection{Caveats	}
 The $\Mcl^{-2}$ scaling of cluster masses [Eq.~(\ref{eq:Mcl_distribution})] is based on present-day observations. A larger number of heavier clusters might have been present (and subsequently disrupted) at larger redshifts. This could increase the fraction of binaries retained in the cluster, and hence the number of 2g mergers. 
Moreover, we fixed $r_h$ as a function of $\Mcl$ [Eq.~(\ref{eq:rh})] based on fits from Ref.~\cite{Antonini:2016gqe}. However, a wide distribution of $\rh$ is observed for a given $\Mcl$. This means that even for small $\Mcl$, one can find smaller $\rh$ (and hence larger $\vesc$) than assumed here. Therefore, 2g mergers could occur efficiently in many clusters smaller than those discussed here.
We also assumed that all clusters have the same distribution of BHs given by $\alpha$ and $\mmax$. More realistically, the mass distribution of BHs should be a function of metallicity, which in turn has a complicated correlation with $\Mcl$, the mass of the host galaxy, etc~\cite{2018MNRAS.480.2343C}.  These complications were ignored in our simple model, and they are beyond the scope of the present work.}

\section{Filling the mass and spin gaps by hierarchical mergers}
\label{sec:massspin}

We now discuss two key features of the observed merger distribution that can help to identify the origin of BBH mergers: the mass gap and the spin gap.

\subsection{The mass gap}

Theoretical studies suggest the existence of a gap in the BH mass function of 1g mergers above $\mmax \sim 45M$~\cite{Farmer:2019jed}  due to PISN and PPISN~\cite{Woosley:2016hmi}. The distribution of BBH mergers detected in O1 and O2 already hints at the possible existence of an upper bound of $\sim 40\Msun$ on the component BH masses~\cite{LIGOScientific:2018jsj}.  %
The detection of BH binaries with component masses in the mass gap could be evidence that repeated mergers are at play.

In our model, only 2g BHs can have masses above $m_{\rm max}$.  We thus define the number of mergers in the mass gap as
\be
\Nmass\equiv N_{1\text{g}+2\text{g}}(m_1>\mmax)+N_{2\text{g}+2\text{g}}(m_1>\mmax)\,.
\ee
The efficiency of the cluster model at populating the mass gap is given by
\bea
\lmass \equiv \f{\Nmass}{\Ncl}\,.\label{lmass}
\eea

The top panel of Fig.~\ref{fig:hist} shows the distribution of primary masses for $\chimax=0.1,\,0.2$. \new{In both cases,} around $25\%$ of 2g mergers lie in the mass gap. 

\subsection{The spin gap}

While core collapse might leave behind slowly rotating BHs \cite{Fuller:2019sxi},
BBH mergers produce remnants with a spin distribution peaked at $\chi\sim 0.7$ \cite{Berti:2008af,Gerosa:2017kvu,Fishbach:2017dwv}. This is the second smoking-gun signature of 2g mergers, as shown in the bottom panel of Fig.~\ref{fig:hist} for $\chimax=0.1$. The effective spin distributions of the 2g+1g and 2g+2g populations is broader compared to the 1g+1g case, with events leaking in the region where $|\chieff|>\chimax$. This region is the spin gap.

Much like evading the PISN/PPISN constraint is a prerogative of repeated mergers, we find that populating the spin gap is also a strong indication of 2g events. The 2g+1g and 2g+2g populations are, collectively, well distinct from the 1g+1g binaries. In particular, 2g events constitute  only $\sim 1\%$
of mergers with effective spin outside the gap ($|\chieff|<\chimax$ ). 
This assumption is solid as we change $\chimax$: we find that the fraction of 2g events outside the spin gap peaks at $2\%$ for $\chimax=0.24$.

The number of events in the spin gap is defined as %
\be
\Nspin\equiv N_{1\text{g}+2\text{g}}(|\chieff|>\chimax)+N_{2\text{g}+2\text{g}}(|\chieff|>\chimax)\, ,
\ee
while
\bea
\lspin \equiv \f{\Nspin}{\Ncl}\label{lspin}
\eea
is the cluster efficiency at populating this region. In the bottom panel of Fig.~\ref{fig:hist}, for example, $74\%$ of all 2g mergers lie in the spin gap when $\chimax=0.1$. This number reduces to $50\%$ for $\chimax=0.2$.

\begin{figure}[t]
  \includegraphics[width=\columnwidth]{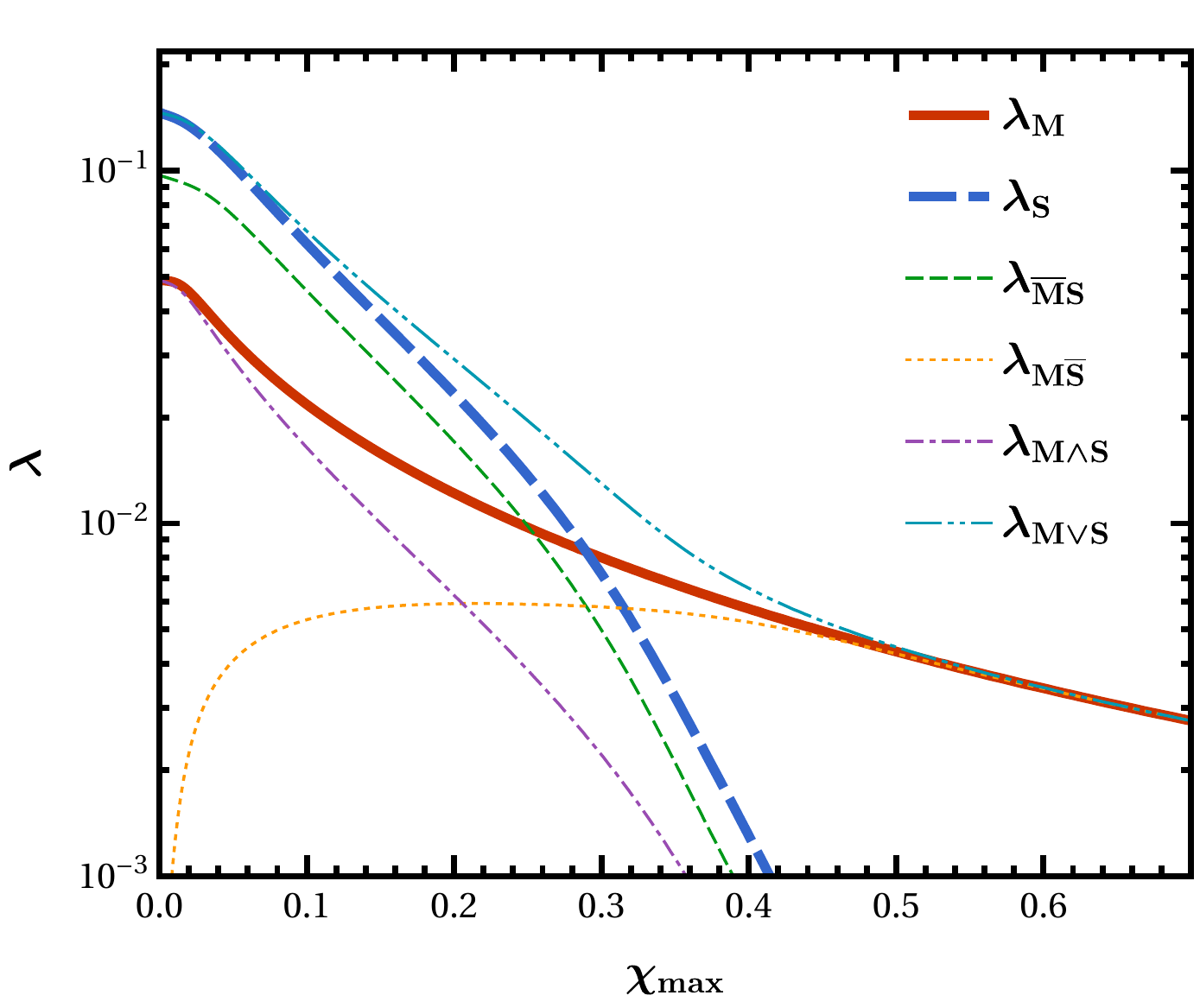}
 \caption{Fraction of events that lie in one and/or both gaps. The total contributions to the mass  ($\lmass$) and spin ($\lspin$) gap are given by the thick solid red and thick dashed blue curves, respectively. The other curves indicate contributions from binaries that are in one gap but not in the other one ($\lambda_{\rm M\overline{S}}$ and $\lambda_{\rm \overline{M}{S}}$), in both gaps  ($\lambda_{\rm M\land S}$), or in either of the two gaps ($\lambda_{\rm M\lor S}$).}
\label{fig:lambda5}
\end{figure}

\subsection{Gap efficiencies}
\label{gapefficienties}

The maximum 1g spin $\chimax$ is the main parameter that determines the efficiencies $\lmass$ and $\lspin$, which are shown as thick lines in Fig.~\ref{fig:lambda5}. For $\chimax=10^{-2}$, about $5\%$ ($14\%$) of all mergers lie in the mass (spin) gap.
These are conservative upper limits: if $\chimax$ is increased, merger products receive larger and larger GW kicks, and the number of 2g mergers decreases drastically. For $\chimax=0.5$, only $0.4\%$ of mergers lie in the mass gap. The effect is even more severe for the spin gap, which remains nearly empty ($\simeq 0.01\%$ of the events).

\begin{figure}[t]
  \includegraphics[width=\columnwidth]{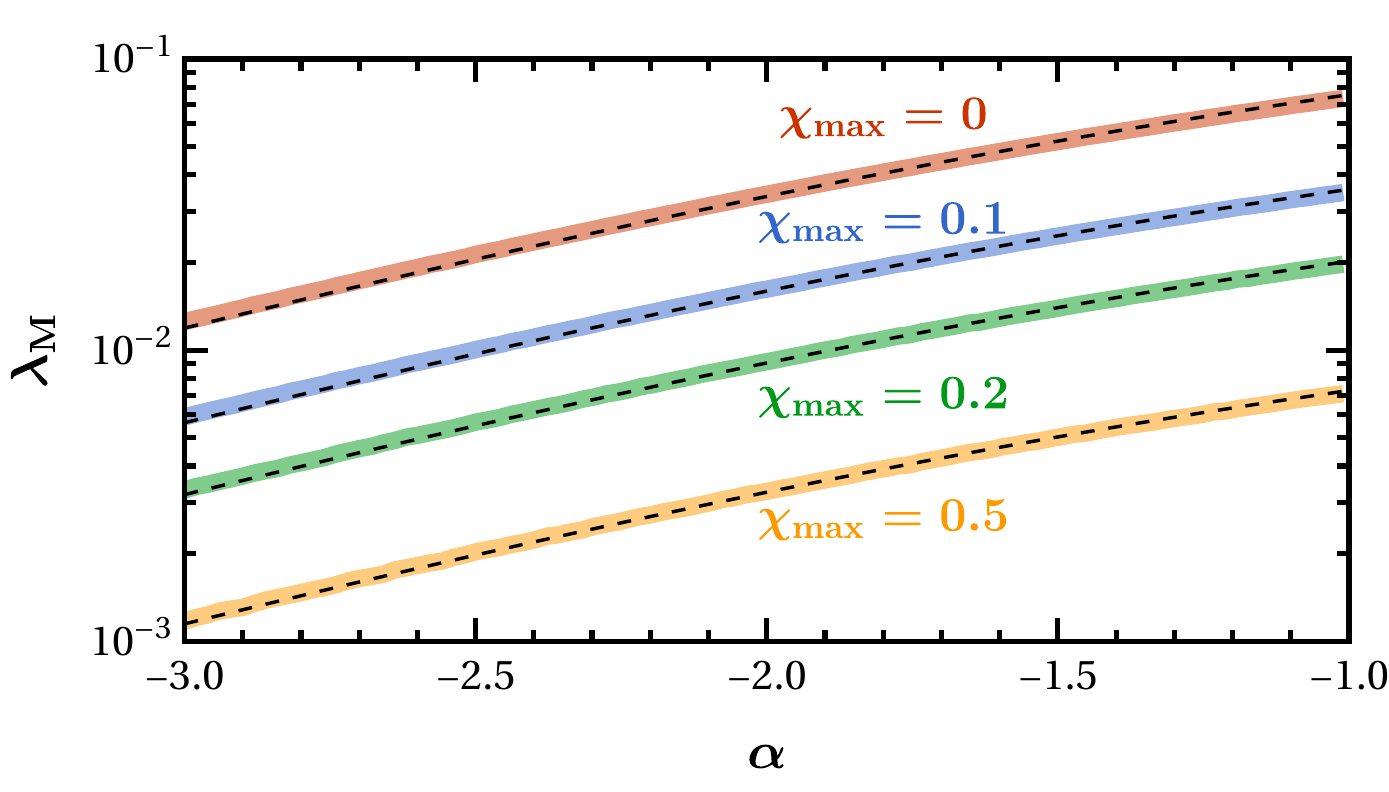}
  \includegraphics[width=\columnwidth]{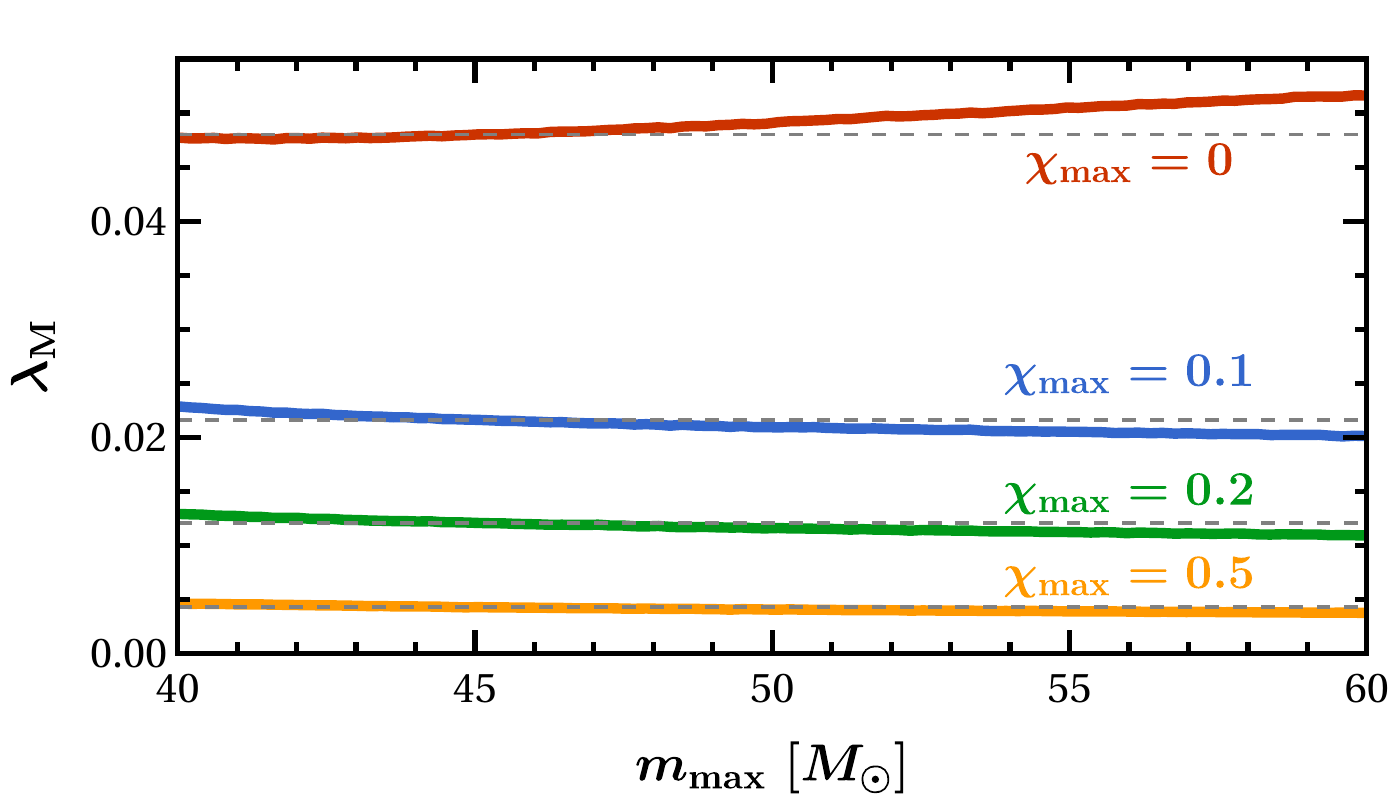}
  \caption{Mass-gap efficiency $\lmass$ as a function of the mass spectral index $\alpha$ (top panel) and the gap edge $\mmax$ (bottom panel). The largest 1g spin $\chimax$ is varied from $0$ to $0.5$ (top to bottom in each panel). In the top panel, black dashed lines show the approximate dependence from Eq.~(\ref{eq:scaledl})  with $\alpha=-1.6$. In the bottom panel, black dashed lines represent our default value $\mmax=45 \Msun$.}\label{fig:ma_dep}
\end{figure}

The spin efficiency is largely independent of both  $\mmax$ and $\alpha$. The mass efficiency, on the other hand, depends on $\alpha$ and, to a lesser extent, on $\mmax$. These trends are illustrated in Fig.~\ref{fig:ma_dep}, which can be understood by a simplified model as follows.

Let us assume that the component masses of the  population retained after merger follow some power law $p_{\rm 1g}(m) \propto m^{\alpha'}$ for $m \in [m_{\rm min,is}, \mmax]$. Let us also neglect energy dissipation during merger. The primary masses of 2g events will be distributed according to%
\be
p_{\rm 2g}(m)=\f{1}{2} \f{1+\alpha' }{\mmax^{1+\alpha' }-m_{\rm min,is}^{1+\alpha' }} \left(\f{m}{2}\right)^{\alpha'}\, ,
\ee
where $m \in [2 m_{\rm min,is}, 2 \mmax]$. The probability that an event lies in the mass gap, i.e. $m>\mmax$, is given by
\bea\label{eq:lambdatoy}
{\Lambda}(\alpha') &=& \int_{\mmax}^{2\mmax} p_{\rm 2g}(m) \dd m =
\f{2^{-(\alpha'+1)}-1}{Q^{-(\alpha'+1)}-1} \,,
\eea
where
\be
Q \equiv \f{\mmax}{m_{\rm min,is}}\sim 7.5 \,.
\ee
We can approximate the dependence of $\lmass$ on $\alpha$ by rescaling

\be\label{eq:scaledl}
\lmass(\chimax,\alpha) = \f{{\Lambda}(\alpha)}{{\Lambda}(\alpha')} \lmass(\chimax,\alpha')\,.
\ee
The scaled value $\lambda(\alpha=-1.6)$ is shown  in the top panel of Fig.~\ref{fig:ma_dep} with black dashed lines. Our analytical approximation closely follows the estimate provided by the full cluster model. This agreement is somewhat surprising, because the primary components of retained BHs are not distributed according to a power law at low masses (cf. the right panel of Fig.~\ref{fig:Mcl}).  

The bottom panel of Fig.~\ref{fig:ma_dep} shows that $\lmass$ is very mildly dependent on $\mmax$. For example, for $\chimax=0.1$, $\lmass$ changes from $2.5\%$ at $\mmax=40\Msun$ to $2\%$ at $\mmax=60\Msun$.

\begin{figure}[t]
  \includegraphics[width=\columnwidth]{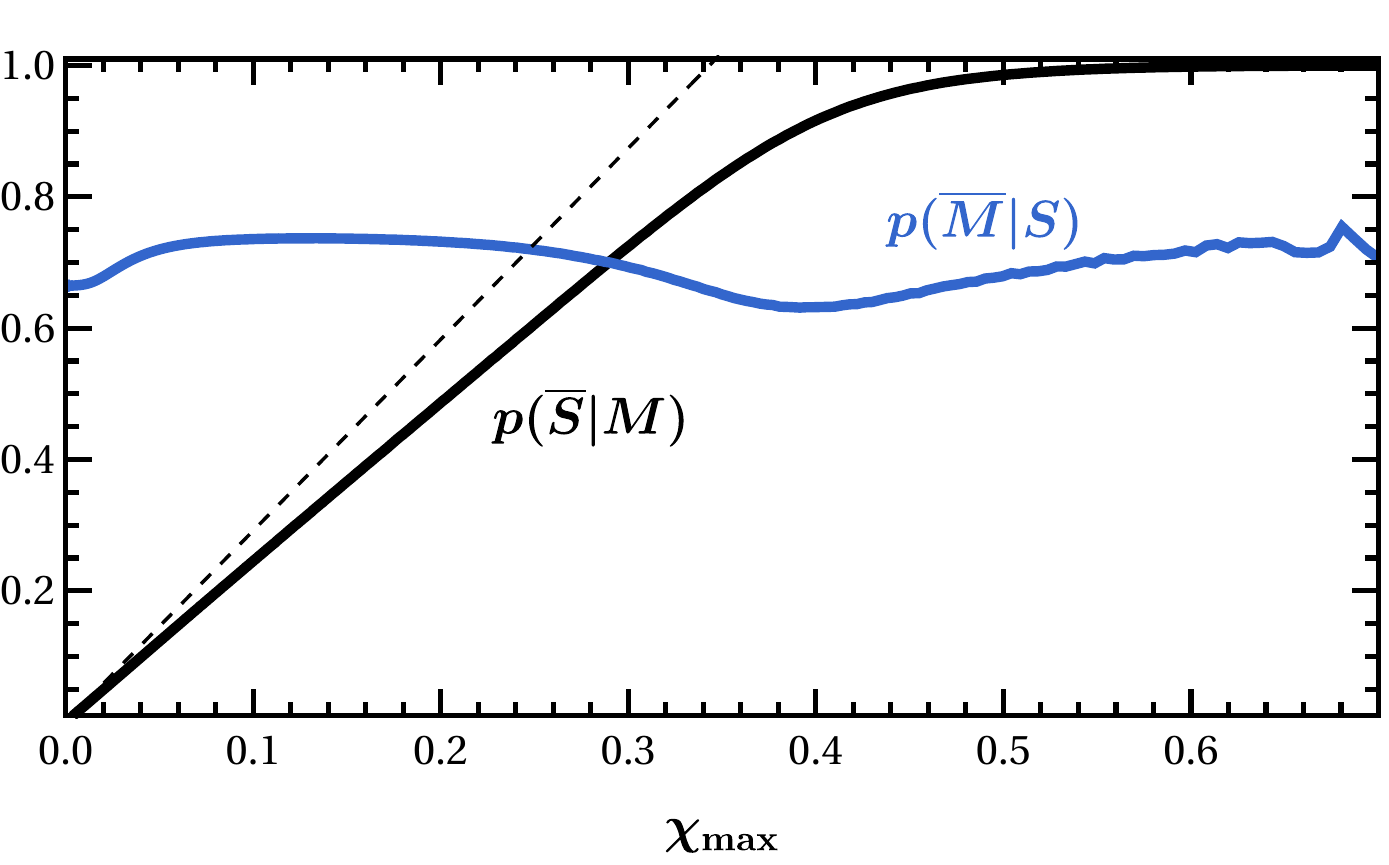}
  \caption{Probability of an event being in only one of the two gaps. The blue (black) curve shows the probability that a binary lies in the  the spin (mass) gap but not in the mass (spin) gap. The dashed black line corresponds to the approximation $\chimax/0.34$ (see text).  } 
\label{fig:spin_or_mass}
\end{figure}

\subsection{One or both gaps?}
\label{sec:onegap}

Although we argue that both gaps are smoking-gun signatures of hierarchical mergers, only a subset of binaries will have both large masses and large spins. Other sources will lie in one of the gaps but not the other.

The various contributions are shown in Fig.~\ref{fig:lambda5}. In particular:
\begin{itemize}
\item $\lspin$ and $\lmass$ are the spin- and mass-gap efficiencies introduced above in Eqs.~(\ref{lmass}) and (\ref{lspin});

\item $\lambda_{\rm M \land S}$ is the fraction of events that lie in {\em both} the mass gap and the spin gap;

\item $\lambda_{\rm M \lor S}$ is the fraction of events that lie in {\em either} the mass gap or the spin gap;

\item $\lambda_{\rm M\overline{S}}$ is the fraction of events that lie in the mass gap, but not in the spin gap; and  

\item $\lambda_{\rm \overline{M}S}$ is the fraction of events that lie in the spin gap, but not in the mass gap. 
\end{itemize}

For $\chimax\sim 0$, the spin gap occupies a large portion of the parameter space, which implies that all binaries in the mass gap must  also be in the spin gap: $\lmass\to \lambda_{\rm M \land S}$. The opposite is true for $\chimax\gtrsim 0.7$: the spin gap shrinks and, consequently, $\lmass\to \lambda_{M\overline{S}}\sim \lambda_{\rm M \lor S}$. 
A future event with large mass but small effective spin ($M\overline{S}$) can be explained by our model only if $\chimax$ is sufficiently large. If BHs are indeed born with negligible spins $\chimax\sim 10^{-2}$, we find that \emph{a mass gap event should also be in the spin gap.} This is an important feature of our model, which can potentially allow us to disentangle the contribution to the mass gap provided by hierarchical mergers (as considered here) from other mechanisms.

We can similarly define the probability of producing events in one of the two gaps but not in the other:
\bea
p(\overline{S}|M)  &=&  
 \f{\lambda_{\rm M\overline{S}}}{\lmass} = 1-\f{\lambda_{\rm M \land S}}{\lmass}
\\
p(\overline{M}|S)  &=& %
\f{\lambda_{\rm \overline{M}S}}{\lspin} = 1-\f{\lambda_{\rm M\land S}}{\lspin}\,.
\eea
These are shown in Fig.~\ref{fig:spin_or_mass}. Events in the spin gap have a $\sim67\%$ probability of not being in the mass gap (blue curve) at $\alpha=-1.6$. Notably, this probability is almost independent of $\chimax$. On the other hand, the probability of observing a mass gap event outside the spin gap grows from 0 to 1 as $\chimax$ increases. 

More specifically, we find a linear behavior  $p(\overline{S}|M)\propto \chimax$ for $\chimax\to 0$. As we show in Appendix~\ref{app:chieff}, the $\chieff$ distribution of 2g+1g mergers (which form the bulk of 2g mergers) is roughly uniform for $\chieff\in [-\chi_f/2, \chi_f/2]$, where $\chi_f\simeq0.68$ is the most probable remnant spin (cf. Fig.~\ref{fig:hist}). To a first approximation and for small $\chimax$, the probability of a mass-gap event not lying in the spin gap for 2g+1g mergers can be approximated by $\chimax/(\chi_f/2)\simeq \chimax/0.34$ (black dashed line in Fig.~\ref{fig:spin_or_mass}). 

\new{In this work we have ignored the relatively rare possibility of 3g mergers, which happen when the remnant of a 2g merger is also retained in the cluster. Since 2g BHs have large spins $\sim 0.7$, they receive large merger recoils. Only clusters with very high escape speed can successfully retain a meaningful fraction of 3g mergers~\cite{Gerosa:2019zmo}. For example, at $\chimax\approx0$, when 2g events account for $>10\%$ of all cluster events, only $\sim \mathcal{O}(0.01\%)$  events are 3g.}

\section{Inference with mass and spin gaps}
\label{sec:first_gen}

The effective spin has long been proposed as a tool to infer the fraction $f$ of BHs formed in clusters. The orientations of BHs formed in clusters should be isotropically distributed, leading to a symmetric $\chieff$ distribution centered at $\chieff=0$, while field binaries should be preferentially aligned, leading to a distribution skewed towards positive values of $\chieff$~\cite{Rodriguez:2016vmx,Farr:2017gtv,Wysocki:2017isg,Gerosa:2018wbw,Ng:2018neg}. However this argument fails if 1g BHs are all born with small spins, because in that case the effective spin  $\chieff\sim0$ irrespective of the spin orientations. Indeed, as we show below, if we only focus on 1g+1g mergers the error on the mixing fraction $f$ scales as $\delta f \propto {1}/{\chimax}$.

In this paper we argue that the mass and the spin gap can provide a powerful alternative which, crucially, remains viable also for small  BH spins at birth.

\subsection{Measuring the mixing fraction and $\chimax$ with 1g mergers}
\label{1gerrors}

As we argued earlier, the efficiencies $\lmass$  and $\lspin$ depends on $\chimax$. We first need to estimate the accuracy with which GW detectors can measure $\chimax$ using only 1g BHs.

To simplify the notation, let us introduce a ``normalized effective spin''
\be
\chihat\equiv\f{\chieff}{\chimax}= \f{\chi_1 \cos\theta_1+q \chi_2\cos\theta_2}{\chimax( 1+q)} \in [-1,1]\,.
\label{rescaledchihat}
\ee
We distribute the spin angles $\theta_1$ and $\theta_2$ uniformly in the cosine between $0$ and $\theta_{\rm max}$. For the cluster model, we set $\theta_{\rm max}=\pi$, such that the distribution is isotropic. For the field binaries, perfect alignment would imply $\theta_{\rm max}=0$, but nonvanishing values are predicted in more realistic models (e.g. \cite{Gerosa:2013laa,Gerosa:2018wbw}). Hereafter, $\theta_{\rm max}$ refers to the largest spin tilt of the field binaries.

If $\chihat$ has PDF $\phat({\chihat})$, the PDF of $\chieff$ can be recovered easily as
\be
p(\chieff | \chimax)= \f{\phat\left({\chihat}\right)}{\chimax}\,.
\label{pcondchieffchimax}
\ee
\begin{figure}[t]
  \includegraphics[width=\columnwidth]{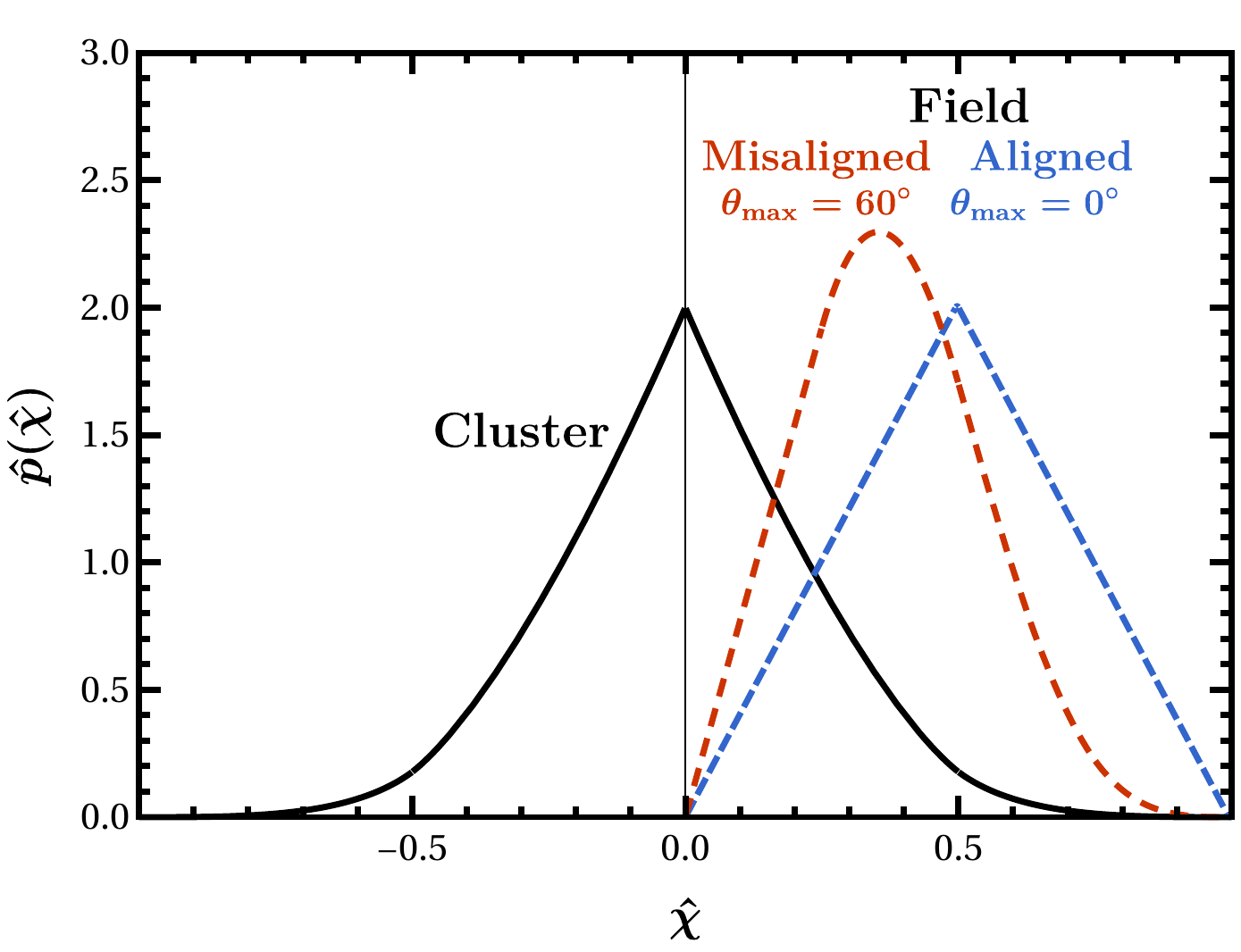}
 \caption{PDF $\pcl(\chihat)$ for cluster binaries from Eq.~(\ref{eq:pcluster}) (solid black line) and field binaries $\pf(\chihat)$ from Eq.~(\ref{eq:pfield}) (dashed lines). For the field binaries, we assume either $\theta_{\rm max}=0\degree$ (blue) or $\theta_{\rm max}=60\degree$ (red). %
 } 
\label{fig:phat}
\end{figure}
\begin{figure*}[t]
  \includegraphics[width=\columnwidth]{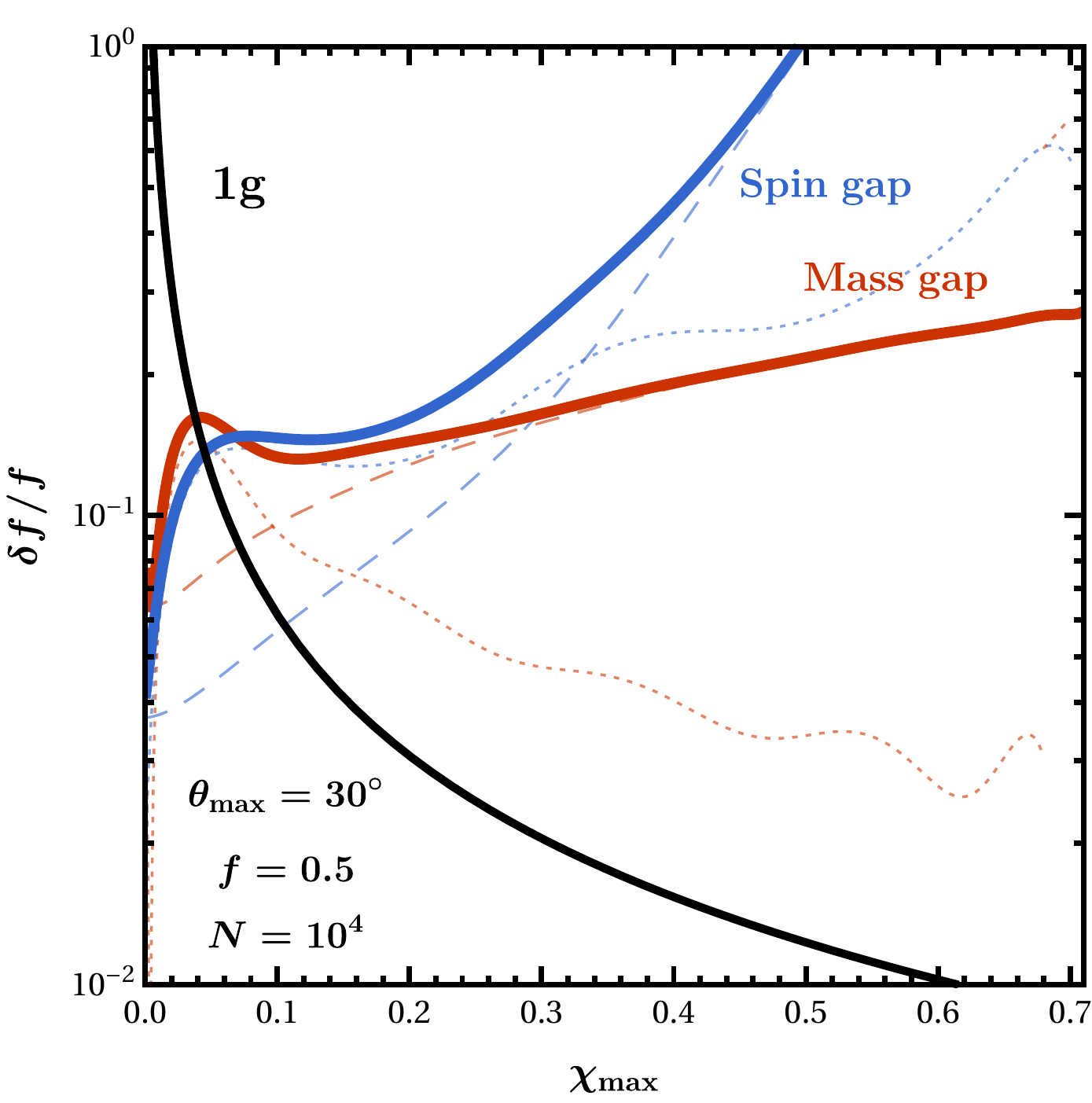}
  \includegraphics[width=\columnwidth]{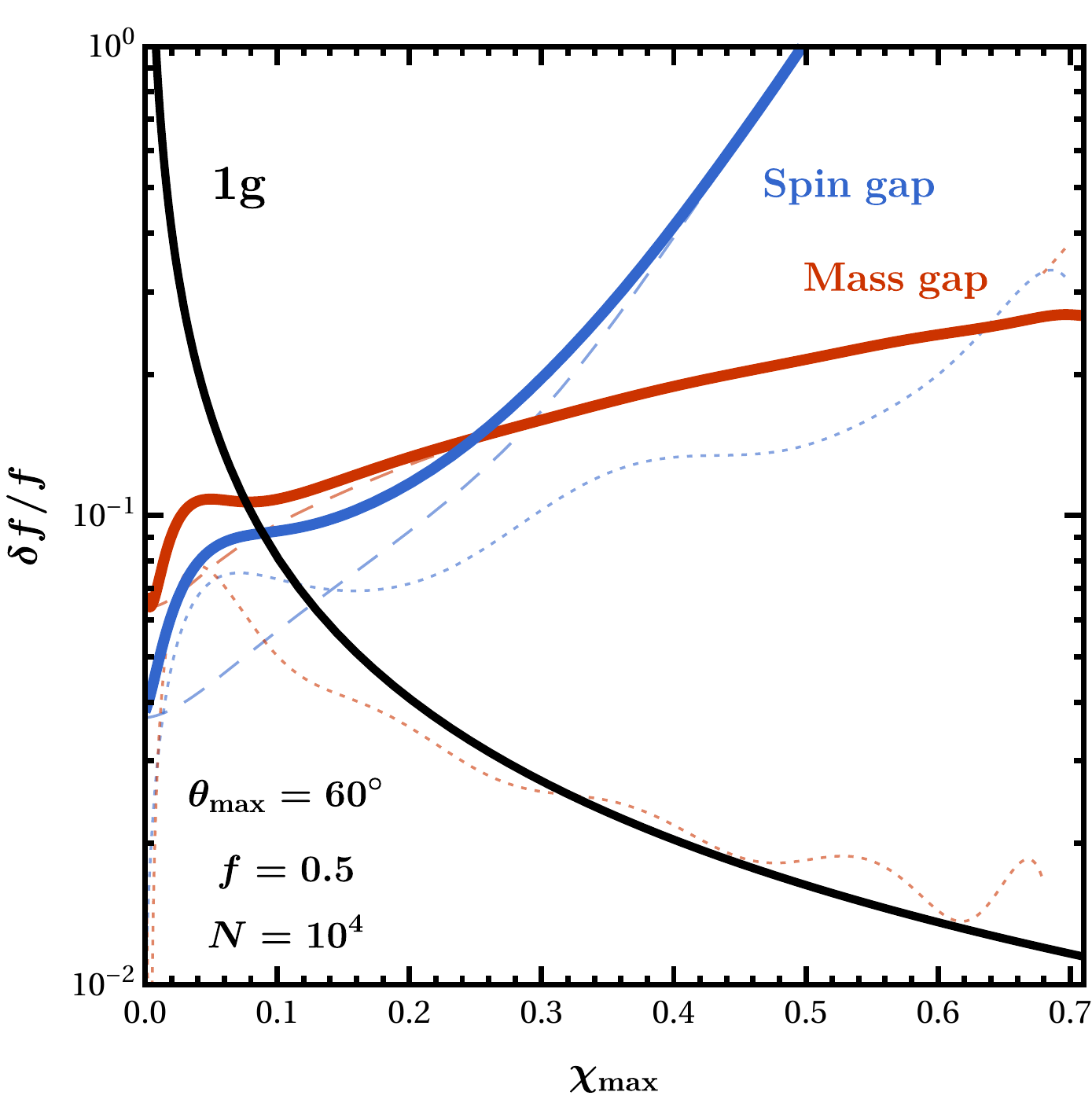}
 \caption{Relative error $\delta f/f$ on the fraction of dynamical mergers as as a function of $\chi_{\rm max}$ considering either only 1g mergers (black), the mass gap (red), or the spin gap (blue). We assume a catalog of $N=10^4$ observations, a mixing fraction $f=0.5$, and the largest misalignment angles for field binaries $\theta_{\rm max}=30\degree$ (left) and $60\degree$ (right). The contributions due to Poisson counting errors and efficiency uncertainties are marked with dashed and dotted lines, respectively.}
\label{fig:delta_f}
\end{figure*}

The derivation of analytical approximations for $\phat \left({\chihat}\right)$ for cluster and field binaries is presented in Appendix~\ref{app:chieff}.  %
The main result consists of the PDFs for the cluster model $\hat{p}_{\rm cluster}(\chihat)$ [Eq.~(\ref{eq:pcluster})] and that of the field formation channel $\hat{p}_{\rm field}(\chihat)$ [Eq.~(\ref{eq:pfield})], shown here in Fig.~\ref{fig:phat}.

If we denote by $f$ the mixing fraction between the two channels, the total PDF is given by 
\be
\phat(\chihat)= f\,\pcl(\chihat) + (1-f)\,\pf(\chihat)\,.
\label{eq:f}
\ee

Suppose we have  $N_{\rm 1g}$ effective spin measurements $\{\chieff^i\}$. The log-likelihood of this sample can be written as
\be\label{eq:L0}
\sL\equiv \sL(f,\chimax|\{\chieff^i\})=\sum_i^{N_{\rm 1g}}\log \f{\phat(\chihat)}{\chimax}\,.
\ee
The quantities $f$ and $\chimax$ can be estimated from the observed $\{\chieff^i\}$ by maximizing the likelihood. The variance of the estimator quantifies the associated uncertainties.%

\subsubsection{Mixing fraction errors}
\label{sec:deltaf_1g}

LIGO measures $\chieff$ better than any other spin parameter, but still with some errors. %
The uncertainties $\delta\chieff^i$ in the $i$-th event affect the estimate of $f$ via
\be\label{eq:delta_f}
(\delta f)^2= \sum_i^{N_{\rm 1g}} \left(\f{\pa f}{\pa \chieff^i}\delta\chieff^i\right)^2\,,
\ee
where
\be
\f{\partial f}{\partial \chieff^i} =  -\f{\pa^2 \sL/ \pa \chieff^i\pa f }{ \pa^2 \sL/\pa f^2}=- \f{\sL_{f\chieff^i}}{\sL_{ff}}
\ee
and 
\begin{align}
\sL_{f\chieff^i}\!\!&=\!\!\sum_i\frac{\chihat_i[  \pcl(\chihat_i ) \pf'(\chihat_i )-  \pf(\chihat_i ) \pcl'(\chihat_i )]}{\chimax\ \phat({\chihat_i})^2}.
\nn
\\
\sL_{ff}&= -\sum_i\left[\f{\pcl(\chihat_i )-\pf(\chihat_i )}{\phat({\chihat_i})}\right]^2\,,
\label{eq:den_f}
\end{align}

Instead of evaluating these quantities as Monte Carlo sums, we consider the integrated expectation value
\be\label{eq:delta_chi_avg}
\langle g(\chihat)\rangle \equiv \f{1}{{N}}\sum_{i} g(\chihat_i) \to  \int_{-1}^1 g(\chihat)\ \phat(\chihat)\ \dd \chihat\,.
\ee
We also express $\delta\chieff^i$ as
\be
\delta\chieff^i =\f{\sigma_{\rm LIGO}}{\rho_i}\,,
\ee
where the signal-to-noise ratio (SNR)  $\rho_i$ is drawn from the distribution $p(\rho)\propto\rho ^{-4}$~\cite{Schutz:2011tw,Chen:2014yla} 
 in the range $[8,\infty)$, and  $\sigma_{\rm LIGO}=\langle\rho\ \delta\chieff\rangle \simeq 1.4$ is the median error $\delta\chieff$ from LIGO/Virgo observations scaled by the SNR~\cite{LIGOScientific:2018mvr}. %
Ignoring the relatively weak SNR dependence on the effective spins~\cite{Dominik:2014yma,Ng:2018neg,Gerosa:2018wbw} and marginalizing over $\delta\chieff$, a Monte Carlo sum like that in Eq.~(\ref{eq:delta_f}) can be approximated as
\bea
 \f{1}{N}\sum_i^N(g(\chihat_i) \delta\chieff^i)^2  &=&\langle g(\chihat)^2 \rangle \int^\infty_{\rho_{\rm thr}} \frac{\langle\rho\ \delta\chieff\rangle^2}{\rho ^2}  p(\rho) \dd \rho \nn\\
 &=& \langle g(\chihat)^2 \rangle (\chierr)^2 \,,
\label{MCkind}\eea
where
\be
\chierr = \sqrt{\f{3}{5}} \frac{\sigma_{\rm LIGO}}{\rho_{\rm thr}}\simeq 0.136\,.
\label{eq:chierr}
\ee

By combining Eqs.~(\ref{eq:delta_f}-\ref{MCkind}) one gets
\be\label{eq:delta_f_1g}
\delta f= \f{{\mathcal F}_f(f,\theta_{\rm max}) }{\chimax \sqrt{N_{\rm 1g}}}\,,
\ee
where
\be
{\mathcal F}_f(f,\theta_{\rm max})^2=
\left(\chierr\right)^2\f{\Big\langle \left(\chihat\frac{  \pcl \pf'-  \pf\pcl'}{\phat^2} \right)^2 \Big\rangle}{\Big\langle \left(\f{\pcl-\pf}{\phat}\right)^2 \Big\rangle^2}
\ee
and we omitted the arguments of $\phat(\chihat)$, $\pcl(\chihat)$ and $\pcl(\chihat)$, for clarity. The function ${\mathcal F}_f(f,\theta_{\rm max})$ varies only mildly with $f$ and increases slowly with the largest misalignment angle of field binaries $\theta_{\rm max}$. For $\theta_{\rm max}<60\degree$, one has ${\mathcal F}_f(f,\theta_{\rm max}) \sim {\mathcal O}(0.1)$.
Equation~(\ref{eq:delta_f_1g}) returns $\delta f \propto \chimax^{-1}$, as expected (see the solid black lines in Fig.~\ref{fig:delta_f}): one cannot rely on the spin orientations to measure the mixing fraction if BH spins are too low. If $\chimax\sim 0.01$~\cite{Fuller:2019sxi}, %
 one would need $\gtrsim\mathcal{O}(10^5)$ 1g detections to achieve errors on the mixing fraction $\delta f\sim 0.1$. For larger values of $\chimax$, we can use 1g events to measure $f$ quite accurately. For example, if $\chimax=0.5$ we can achieve an error $\delta f\lesssim 0.1$ with only $\sim 100$ events.

\subsubsection{Errors on $\chimax$}
\label{sec:dchimax}

We estimate $\delta\chimax$ due to errors on the individual $\chieff^i$ measurements by error propagation:
\be\label{eq:delta_chimax}
(\delta\chimax)^2= \sum_i^{N_{\rm 1g}} \left(\f{\partial \chimax}{\partial \chieff^i}\delta\chieff^i\right)^2\,,
\ee
where using the same notation as above we get
\be
\f{\partial \chimax}{\partial \chieff^i}= -\f{\pa^2 \sL/ \pa \chieff^i\pa \chimax }{ \pa^2 \sL/\pa \chimax^2} =- \f{\sL_{\chimax\chieff^i}}{\sL_{\chimax\chimax}}\,,
\ee
with%
\begin{align}
&\sL_{\chimax\chieff^i} \!= -\frac{1}{\chimax^2} \sum_i \left[\chihat_i \log'' \phat(\chihat_i)+\log'\phat(\chihat_i)\right], \nn\\
&\sL_{\chimax\chimax} = \frac{1}{\chimax^2}\sum_i \!\left[ 1\!+\!2 \chihat_i\log'\phat(\chihat_i)\!+\! \chihat_i^2 \log''\phat(\chihat_i)\right].
\end{align}
Combining these results and replacing the sum by an integral as in Eq.~(\ref{eq:delta_chi_avg}), Eq.~(\ref{eq:delta_chimax}) yields
\be
\delta \chimax= \f{{\mathcal F}_{\chimax}(f,\theta_{\rm max}) }{\sqrt{N_{\rm 1g}}}\,,
\ee
where
\be
{\mathcal F}_{\chimax}(f,\theta_{\rm max})^2=
\left(\chierr\right)^2\f{\langle \left( \chihat \log''\phat+\log'\phat\right)^2 \rangle}{\langle 1+2 \chihat \log'\phat+ \chihat^2 \log''\phat  \rangle^2}\,.
\ee
Here, again, we have suppressed the argument of $\phat(\chihat)$.

The function ${\mathcal F}_{\chimax}(f,\theta_{\rm max})$ presents a mild dependence on $f$ for nonzero misalignment angles, but it diverges as $\theta_{\rm max} \to 0$.  Mathematically, the reason for the divergence is that, in the limit $\theta_{\rm max} \to 0$, $\log'' \phat(\chihat)$ diverges as $\chihat\to 1$ in our simple analytical model. The divergence may not occur in a more accurate Bayesian inference analysis. In the limit $f\to 1$, when all events come from clusters [Eq.~(\ref{eq:f})] and $\theta_{\rm max}$ is irrelevant, we have ${\mathcal F}_{\chimax}(f,\theta_{\rm max})\simeq1$.

Our estimate considers the errors on $\chimax$ and $f$ due only to measurement errors on the $\chieff^i$. Another source of error is the variance of the maximum-likelihood estimator itself (given by the inverse of the Fisher information matrix), which would be present even if all of the $\chieff^i$ were measured perfectly. However these errors are always subdominant with respect to the measurement errors estimated above, because $\chierr$ [Eq.~(\ref{eq:chierr})] is relatively large.

We have also assumed that the maximum likelihood estimator $\chimax^{\rm mle}$ coincides with the true value $\chimax$. This approximation must break down for a finite sample of size ${N}$, and then $\chihat^{\rm mle}=\chieff/\chimax^{\rm mle}$ will not follow the distribution $\phat({\chihat})$.  A more careful error analysis is an interesting topic for future research.

\begin{figure*}[t]
  \includegraphics[width=\columnwidth]{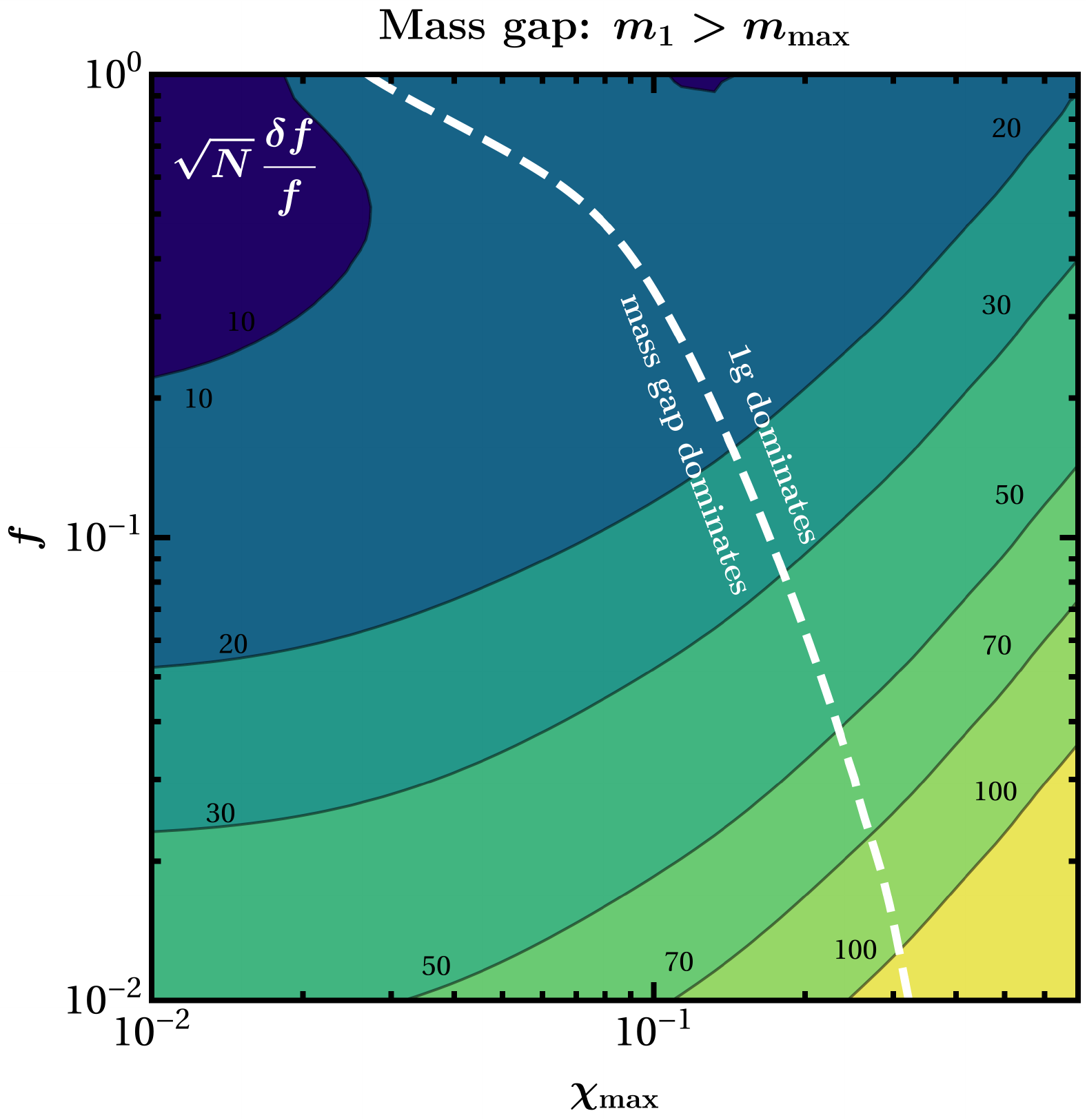}
  \includegraphics[width=\columnwidth]{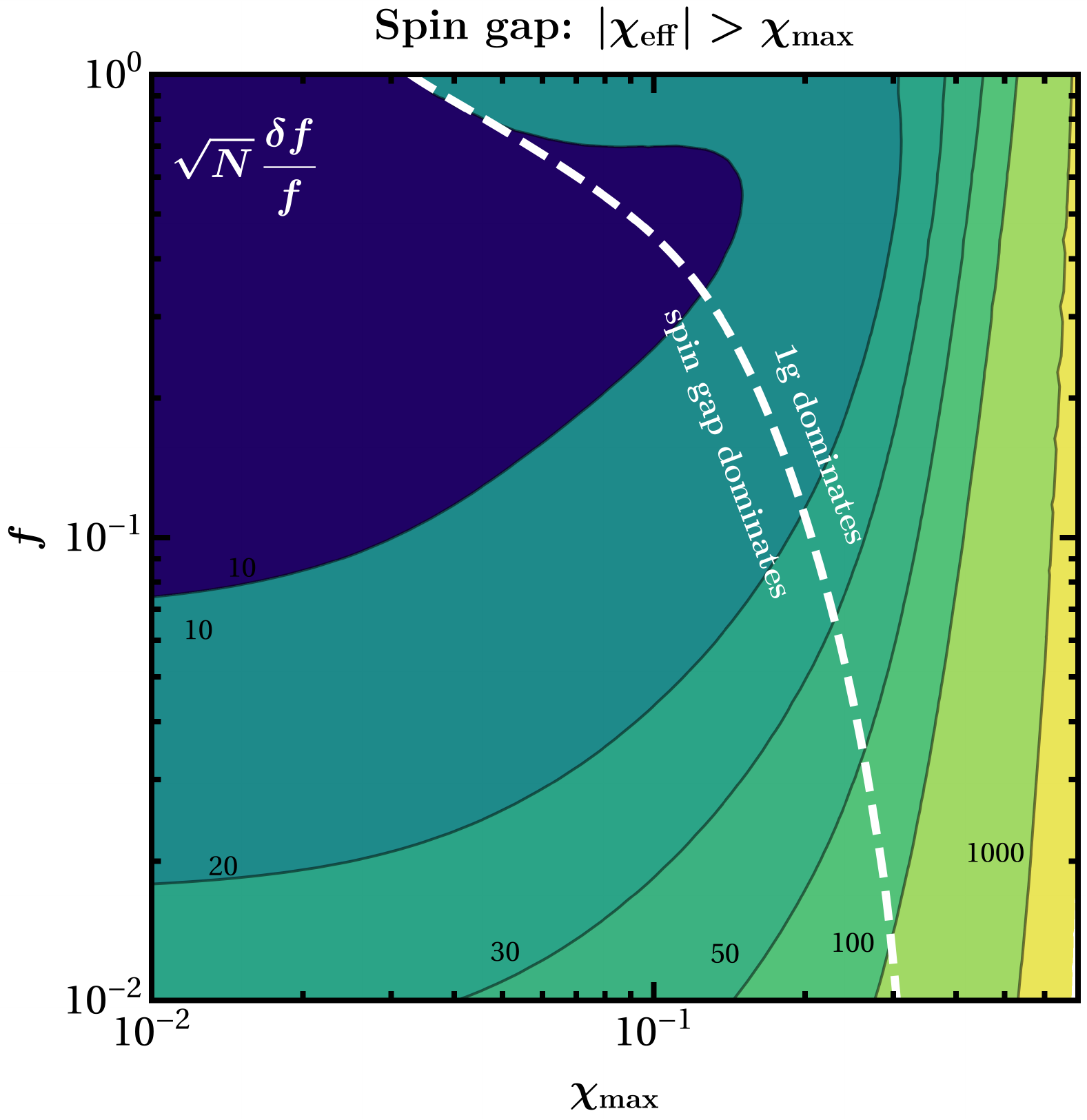}
  \caption{Fractional error on the mixing fraction $\sqrt{N}\delta f/f$ obtained using mass gap (left) and spin gap (right) as a function of $\chimax$ and $f$ assuming $\theta_{\rm max}=60\degree$. The white dashed line marks the location where the error $\delta f$ from 1g  events equals the one obtained with the gaps. In particular, gap (1g) events dominate to the left (right) of the white dashed lines.} 
\label{fig:mass_spin_2d}
\end{figure*}

\subsection{Measuring the mixing fraction with the gaps}
\label{sec:frac}

We finally address the crucial point of this paper: can the mass and spin gaps be used to constrain the mixing fraction between different formation channels?

Suppose we are given a catalog of $N$ observations, which include 1g and 2g events.
As discussed in the introduction, the number of identifiable 2g mergers because of the mass and spin gaps is
\bea
\f{\Nmass}{N}&=&f\ \lmass(\chimax),\,\nn\\
\f{\Nspin}{N}&=&f\ \lspin(\chimax) \,,
\eea
where  $f$ is the fraction of all detections that were produced in clusters. 
The number of 1g events available to infer $\chimax$, as described in Sec.~\ref{1gerrors}, is
\be
N_{\rm 1g} = N \ [1-f \lspin(\chimax)]\,.
\ee

For either the mass gap ($\lambda_{\rm gap}=\lmass$) or the spin gap ($\lambda_{\rm gap}=\lspin$), the uncertainty in measuring $f$ is given by
\be
\left(\f{\delta f}{f}\right)_{\rm gap}^2=\left(\f{ \partial \lambda_{\rm gap}/\partial \chimax}{\lambda_{\rm gap}} \delta\chimax\right)^2+\left(\f{\delta N_{\rm gap}}{N_{\rm gap}}\right)^2\,.
\label{deltafgap}
\ee
The first term on the right hand side represents the uncertainty in measuring the efficiency propagated from $\chimax$.
As we argued in Sec.~\ref{gapefficienties}, the dependencies on the other population parameters (such as $\alpha$ and $\mmax$) are mild and can be neglected.
The second term in Eq.~(\ref{deltafgap}) is the Poisson counting error associated with the number of mass/spin gap events,
\be
\f{\delta N_{\rm gap}}{N_{\rm gap}}=\f{1}{\sqrt{\lambda_{\rm gap} f N}}\, .
\label{countingeq}
\ee

In Fig.~\ref{fig:delta_f} we plot the relative errors in the mixing fraction $\delta f/f$ for both the mass gap (red) and the spin gap (blue) as a function of $\chimax$. Figure~\ref{fig:delta_f} also shows the individual contributions due to efficiency errors and Poisson errors. For illustration we consider $N=10^4$ observations with a mixing fraction of $f=0.5$, and we select two values of the largest misalignment angle for field binaries: $\theta_{\rm max}=30\degree, 60\degree$.

Spin-gap events yield a more accurate measurement of $f$ at small values of $\chimax$. This is because (i) the number of spin-gap events is higher (i.e. $\lspin$ is large) and thus Poisson errors are low, and (ii) the derivative $\partial\lspin/\partial \chimax$ vanishes for $\chimax\to 0$, leading to small efficiency errors. The mass gap is more informative for larger values of $\chimax$, because $\lspin<\lmass$ for $\chimax \gtrsim 0.3$, while $(\partial \lspin/\partial \chimax)/\lspin \,>\, (\partial\lmass/\partial\chimax)/\lmass $ for $\chimax \gtrsim 0.05$.

The ``critical'' value of $\chimax$ at which the mass gap is preferred over the spin gap depends on both $f$ and $\theta_{\rm max}$. %
In particular, the threshold is $\chimax \simeq 0.07 \, (0.25)$ for %
 $f=0.5$ and $\theta_{\rm max}=30\degree \,(60\degree)$.  This is one of the most important findings of this paper: \emph{if BHs are born slowly rotating, spin-gap events are more effective than mass-gap events at pinning down the mixing fraction between formation channels.}

In general, we find that counting  errors dominate for large values of $\chimax$, simply because there are not enough gap events: cf.~Eq.~(\ref{countingeq}). From Fig.~\ref{fig:delta_f}, at $f=0.5$ efficiency errors are important only when $\chimax<0.1\,(0.02)$ for mass-gap events and when $\chimax<0.33\,(0.14)$ for spin-gap events at  $\theta_{\rm max}=30\degree \,(60\degree)$.  The parameter $\theta_{\rm max}$ enters $\delta f/f$ in Eq.~(\ref{deltafgap}) only through the efficiency error [cf. Eq.~(\ref{eq:delta_chimax})]. As a result, only the small-$\chimax$ behavior of the $\delta f/f$ errors is affected by $\theta_{\rm max}$: for example, compare the left and right panels of Fig.~\ref{fig:delta_f}, where $\delta f/f$ is the same for  $\theta_{\rm max}=30\degree$ and $\theta_{\rm max}=60\degree$ at large $\chimax$. %

Besides $\chimax$ and  $f$, the two parameters determining the mass spectrum ($\alpha$ and $\mmax$) mildly affect the number of event, and thus the Poisson errors. For instance, $\lmass$ increases for $|\alpha|<1.6$, which implies that Poisson errors become less relevant. As expected, we find that spin-gap estimates are more robust against changes in $\mmax$ and $\alpha$ compared to mass-gap estimates. This is simply because the mass spectrum parameters have a direct impact on the 1g mass distribution. %

\begin{figure}[t]
  \includegraphics[width=\columnwidth]{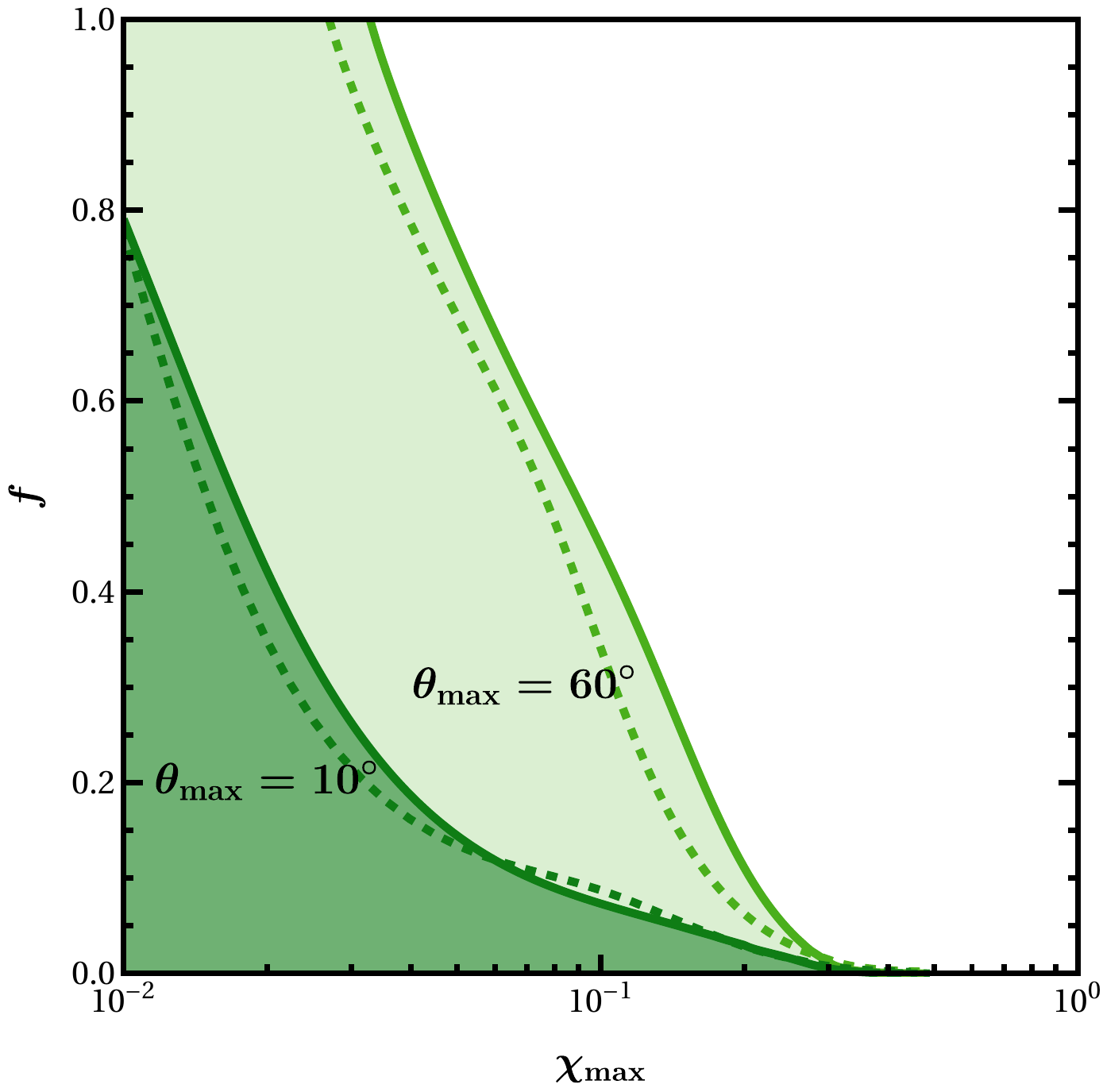}
  \caption{The shaded areas mark values of $f$ and $\chimax$ where gaps provide a more accurate measurement of the mixing fraction $f$ compared to 1g events. Results for the spin (mass)  gap are indicated with solid (dashed) curves. Darker (lighter) regions show results for $\theta_{\rm max}=10^\circ$ ($60^\circ$).}
\label{fig:gap_dominates}
\end{figure}

Figure~\ref{fig:mass_spin_2d} shows contours of ${\delta f}/f\sqrt{N}$ obtained from the mass gap (left panel) and the spin gap (right panel) in the $(\chimax,\,f)$ plane, assuming for concreteness $\theta_{\rm max}=60\degree$. We also plot contours (dashed white lines) where either of the gap measurements and 1g detections achieve the same $\delta f$ accuracy: gaps are better than 1g observations at constraining $f$ to the left of these lines, and worse to the right of these lines. As a rough rule of thumb, {\em the mixing fraction is better constrained through gap measurements when $\chimax\lesssim 0.1$, while the 1g population is more constraining if $\chimax\gtrsim 0.1$.} This is another central result of our work. %

Suppose for concreteness that $\chimax=10^{-2}$, as predicted by Ref.~\cite{Fuller:2019sxi}. Then spin-gap events would allow us to measure the mixing fraction $f$ with an accuracy  of $\delta f =0.1$ with a catalog of  $N\sim 150$ events if $f=0.2$ ($N\sim3000$ events if $f=1$), corresponding to $5$ ($400$) events in the gap. Achieving the same $\delta f =0.1$ accuracy with mass-gap events would require  $N\sim 450$ events if $f=0.2$ ($N\sim4400$ events if $f=1$), corresponding to $4$ ($200$) events with $m_1>\mmax$. This should be compared with the $\gtrsim 10^5$ events needed for measuring $f$ at $\chimax=10^{-2}$ using 1g events only. %

To make the previous rule of thumb more precise, in Fig.~\ref{fig:gap_dominates} we show regions in the $(\chimax,\,f)$ plane where gap events would lead to smaller errors compared to 1g events. This plot confirms that looking at events in the gaps is crucial if BHs are born slowly rotating (i.e., at small $\chimax$) and cluster formation is inefficient (i.e., at small values of $f$).  The region where the gaps are important increases with $\theta_{\rm max}$, because  (as we discussed in Sec.~\ref{1gerrors}) $\delta \chimax$ diverges for small values of $\chimax$.

We summarize our main points as follows:
\begin{itemize}
\item Spin-gap events measure $f$ more accurately than mass-gap events at small $\chimax$.
\item The error $\delta f$ at $\chimax\gtrsim 0$ is dominated by the efficiency error; this is governed by the maximum misalignment angle $\theta_{\rm max}$ in field formation scenarios, and it is only mildly affected by the mass spectrum parameters $\alpha$ and $\mmax$.
\item Poisson errors dominate for large $\chimax$. For spin-gap events the Poisson errors depend only on $\chimax$, while for mass-gap events they are also governed by the mass spectrum parameters $\alpha$ and $\mmax$.
\item Gap events measure $f$ better than 1g events at small $\chimax$, small $f$, and large $\theta_{\rm max}$.
\end{itemize}

\section{Conclusions}
\label{sec:conclusions}

Each event in the growing LIGO/Virgo BBH catalog yields three main ``intrinsic'' observable quantities: the binary component masses and the effective spin. The observed events are likely to come from at least two, and possibly more, formation channels. As the number of observations grows, reserved regions in the intrinsic parameter space (the mass and spin gaps) could allow us to measure the relative contributions of different channels. In this paper we quantified this statement, showing that the mass and spin gaps allow us to efficiently separate the contribution of field binaries from the contribution of binaries formed dynamically in star clusters.

Supernova instabilities \cite{Woosley:2016hmi} and efficient core-envelope interactions \cite{Fuller:2019sxi} imply that massive stars should form BHs with an upper mass ($m\lesssim 45 M_\odot$) and spin ($\chi\lesssim 0.1$) limit. Therefore, our current understanding of field binary evolution hints at the existence of {\em both} a ``mass gap'' and a ``spin gap'' within this formation scenario. Repeated mergers in clusters provide a natural way to evade these constraints (see e.g.~\cite{Christian:2018mjv,Gerosa:2019zmo,Rodriguez:2019huv}).
Assuming to a first approximation that only these two scenarios are at play, observations of BBHs in either the mass gap, the spin gap, or both, would not only imply that those events were formed dynamically, but it would also improve our understanding of the origin of the whole observed population.

The key theoretical input to perform this analysis is the efficiency with which dynamical environments like GCs and NSCs can populate the gap(s). We used a semianalytical model specifically designed to predict the occurrence of repeated (2g) mergers in dense star clusters. The main prediction of our model is that the gap efficiencies are of the order of a few $\%$ and that they are mostly sensitive to $\chi_{\rm max}$, the largest spin magnitude of individual BHs at birth. 

We propose the following broad observational strategy. We can assume that the bulk of the population consists of 1g BHs, which are outside the gaps and can be used to measure $\chimax$ (Sec.~\ref{sec:first_gen}). As shown in Sec.~\ref{sec:massspin}, a measurement of $\chimax$ can be converted into a solid estimate of the cluster efficiencies at populating the mass gap ($\lambda_{\rm M}$) and the spin gap ($\lambda_{\rm S}$). Combined with the measured distribution of the effective spins $\left\{\chieff^i\right\}$, this results in an estimate of the mixing fraction $f$ between cluster and field binaries (Sec.~\ref{sec:frac}).

Many studies in the literature (see e.g.~\cite{Rodriguez:2016vmx,Farr:2017gtv,Gerosa:2018wbw,Sedda:2018nxm,Safarzadeh:2020jsc,Sedda:2020vwo}) devised strategies to infer the mixing fraction $f$ between field and cluster formation channels from the observed distribution of effective spins $\left\{\chieff^i\right\}$. The underlying idea behind most of these studies is that cluster formation predicts a distribution of $\chieff$ which is symmetric about zero, while binaries in the field should show a preference for $\chieff>0$. All strategies that rely on a measurement of the spin orientations, however, are bound to fail if the spin magnitudes are small. Indeed, in Sec.~\ref{sec:first_gen} we show that, within this approach, the accuracy in determining the mixing fraction scales like $\delta f\propto 1/\chimax$.
 
On the other hand, by exploiting the gaps we can estimate the mixing fraction $f$ between different channels {\em even if BH spins at birth are zero.} If anything, the spin gap is larger if the natal spins are small, and outliers can be more easily identified. Indeed, we find that both the mass and spin gaps allow for a better measurement of the mixing fraction (compared to the standard ``$\chieff$ distribution test'')  test as long as $\chimax\lesssim 0.1$. Both observations~\cite{LIGOScientific:2018mvr,LIGOScientific:2018jsj} and theoretical modeling~\cite{Fuller:2019sxi,Belczynski:2020bnq} suggest that this is indeed the case, making our new observational strategy timely and relevant. 

We assumed that the mass and spin gaps can only be populated by repeated mergers in clusters. This is an important caveat of our study. While 2g mergers constitute a well-motivated scenario (see e.g.~\cite{Christian:2018mjv,Gerosa:2019zmo,Rodriguez:2019huv})
which is now being implemented in LIGO/Virgo parameter estimation pipelines~\cite{Doctor:2019ruh}, other astrophysical mechanisms could ``pollute'' the gaps and deteriorate the measurement of $f$. These include include gas accretion~\cite{Roupas:2019dgx}, stellar mergers~\cite{DiCarlo:2019fcq}, Population III stars~\cite{Kinugawa:2015nla,Hartwig:2016nde,Belczynski:2016ieo}, or gravitational lensing\footnote{A gravitational lens with magnification $\mu$ increases the amplitude of a GW event by a factor of $\sqrt{\mu}$~\cite{Ng:2017yiu,Hannuksela:2019kle}.  This magnification reduces the inferred luminosity distance, increasing the apparent source-frame mass and producing ``fake'' mass-gap events. However, the lensing probability is rather small ($\sim 10^{-3}$ according to Ref.~\cite{Ng:2017yiu}). The distribution of magnification factors depends on the lens model, but it roughly scales like $p(\mu)\sim\mu^{-3}$ for $\mu>2$~\cite{1992grle.book.....S,Hilbert:2007ny,2010MNRAS.406.2352L}. In order to contribute significantly to the mass gap, lensed events must be located at high redshifts, have large magnification, and have source-frame masses close enough to the mass gap.  Taking into account the lensing probability, the shape of the mass function and the magnification function, we estimate that the probability of observing a mass-gap event due to lensing is $\sim 10^{-5}$, so it can safely be neglected in the present context.} of 1g events~\cite{Ng:2017yiu,Hannuksela:2019kle}.

An independent reanalysis of data from the first and second LIGO/Virgo observing runs identified at least one BBH event (GW170817A, not to be confused with the famous binary neutron star merger that occurred on the same day) which may have {\em both} high mass and high spin~\cite{Zackay:2019btq}. Based on these properties of the binary, some authors~\cite{Gayathri:2019kop} suggested that GW170817A might have formed in an AGN disk~\cite{Bartos:2016dgn,Stone:2016wzz}.  However, as pointed out in the context of the first candidate mass gap event GW170729~\cite{Fishbach:2019ckx}, it is dangerous to evaluate individual outlier events without reference to the entire population. We postpone a more complete study, including an outlier analysis along the lines of Ref.~\cite{Fishbach:2019ckx}, to future work.

Two predictions of our model are particularly noteworthy, because they could be verified or disproved in the near future:

\begin{enumerate}
\item Future events with large mass but small effective spin ($M\overline{S}$ in our notation) can be explained {\em only if $\chimax$ is sufficiently large}. In other words, we find that \emph{a mass gap event should also be in the spin gap}: cf. Fig.~\ref{fig:lambda5}, and note that Ref.~\cite{Belczynski:2020bnq} recently proposed a similar argument. This is an important feature of our model, that can potentially allow us to disentangle the hierarchical merger contribution to the mass gap considered in this work from other astrophysical mechanisms.

\item If BHs are born slowly rotating, high-spin events are more effective than high-mass events to pin down the mixing fraction between formation channels: the spin gap (which was largely neglected in the literature so far) is actually more discriminating than the mass gap if spins are small, as suggested by astrophysical theory and LIGO/Virgo observations so far (Fig.~\ref{fig:gap_dominates}).
\end{enumerate}

In this exploratory study we have made simplifying assumptions that should be relaxed in the future.

First of all, to keep the analysis general, we focused on the fraction of the total number of observations that end up in the mass or spin gaps. In practice, this fraction will be detector-dependent: third-generation detectors such as the Einstein Telescope or Cosmic Explorer will be more sensitive to low-mass binaries, while current detectors introduce a selection bias that favors large masses.

Secondly, our ability to distinguish between different formation scenarios could improve if we considered not only the number of events in the gap, but also their {\em distribution}. In our strategy we proposed estimating $\chimax$ from the 1g spin distribution and to simply count events with $|\chieff|>\chimax$. However 2g+2g mergers can result in effective spins $\chieff\gtrsim 0.34$, while this is not possible for 2g+1g events (cf. Fig.~\ref{fig:hist}). Third-generation detectors could allow us to infer the spin distribution of spin-gap events, and possibly to measure the relative number of 2g+2g and 2g+1g events.
Similarly, the mass distribution of mass-gap events contains useful information. For large $\chimax$, only very large NSCs could retain post-merger remnants, leading to a steeper mass distribution (cf. Sec.~\ref{hiemer}). This dependence is weak, but mass-gap events should typically have large SNRs, and therefore their mass distribution is easier to measure.

The upcoming release of LIGO/Virgo data from the third observing run O3 will bring us closer to the large-statistics regime of GW astronomy. As we enter this new era, the observational strategy outlined in this paper could lead us to a better estimate of the relative contribution of different formation channels.

\acknowledgments

We thank Alberto Vecchio, Cyril Creque-Sarbinowski, Giacomo Fragione and Mohammadtaher Safarzadeh for discussions.
V.B., E.B., T.H. and K.W.K.W. are supported by NSF Grants No. PHY-1912550 and AST-1841358, NASA ATP Grants No. 17-ATP17-0225 and 19-ATP19-0051, and NSF-XSEDE Grant No. PHY-090003. 
E.B. also acknowledges support from the Amaldi Research Center funded by the MIUR program ``Dipartimento di Eccellenza''~(CUP: B81I18001170001).
D.G. is supported by Leverhulme Trust Grant No. RPG-2019-350. 
The authors would like to acknowledge networking support by the GWverse COST Action CA16104, ``Black holes, gravitational waves and fundamental physics.'' Computational work was performed on the University of Birmingham BlueBEAR cluster, the Athena cluster at HPC Midlands+ funded by EPSRC Grant No. EP/P020232/1, and the Maryland Advanced Research Computing Center (MARCC).

\appendix

\section{Analytical approximations of $\chieff$ probability distributions}
\label{app:chieff}

The goal of Sec.~\ref{sec:first_gen} is to compute errors on the maximum effective spin $\chimax$ and on the mixing fraction $f$ by error propagation, which requires the evaluation of first and second derivatives of the PDFs. In principle this could be done by sampling the distributions and numerically interpolating the results, which however would result in large errors on the derivatives. To overcome this problem, in this Appendix we find analytical expressions for the PDF of $\chihat$ in the two scenarios of interest: cluster and field formation. %

Our starting point is the rescaled effective spin $\chihat$ of Eq.~(\ref{rescaledchihat}).
Since in Eq.~(\ref{eq:pm2}) we set $\beta=6.7\gg1$, 
we can assume $q=1$ for the vast majority of our sources, so that
\be
\chihat\simeq \f{\hat{\chi}_1 \cos\theta_1+ \hat{\chi}_2\cos\theta_2}{2 } \,,
\ee
where $\hat{\chi}_i=\chi_i/\chimax$. We have verified that setting $q=1$ leads to negligible deviations with respect to the PDFs found by using generic values of $q$.

We will repeatedly use the following standard identities from probability theory. %
 If $X$ and $Y$ are two independent, continuous random variables with PDFs $f_X$ and $f_Y$, the PDF of their product $XY$ and of their sum $X+Y$ are
\begin{align}
\label{eq:pdf_prod}
f_{XY}(z)&=\int _{{-\infty }}^{{\infty }}f_{X}\left(x\right)f_{Y}\left(z/x\right){\frac  {1}{|x|}}\,\dd x\,,
\\
\label{eq:pdf_sum}
 f_{X+Y}(z)&=\int _{-\infty }^{\infty }f_X(x)f_Y(z-x)\dd x\,.
\end{align}
The PDF of a generic bijective function $g(X)$ is 
 \be\label{eq:pdf_func}
f_{g(X)}(z) = f_X(g^{-1}(z)) \left\vert \f{\dd g^{-1}(x)}{\dd x}\right\vert_{x=z}\,.
\ee

\subsection{Field binaries}

For field binaries the spins will be nearly aligned, so we draw $\cos\theta_i$ $(i=1,\,2)$ uniformly in the range $[1-\delta,1]$, where $\delta$ is related to the maximum misalignment angle $\theta_{\rm max}$ of each spin by $\theta_{\rm max}=\arccos( 1-\delta)$. %

First we find the distribution of
\be
z_i=\chihat_{i} \cos\theta_{i}\,,
\ee
which is a product of the two uniform distributions
\bea
&p(\cos\theta_i) = {1}/{ 2\delta}, \quad\quad & \cos\theta_i  \in [1-\delta,1]\,,\nn\\
&p(\chihat_i) = 1, \quad\quad & \chihat_i  \in [0,1]\,.
\eea
From Eq.~(\ref{eq:pdf_prod}) one gets
\be
p(z_i)=\begin{cases}
-\log (1-\delta )/\delta & \text{for } 0<z_i \le 1-\delta\,,\\
-\log (z)/\delta & \text{for } 1-\delta \le z_i\le 1\,.\\ 
\end{cases}
\ee
For $q=1$, the distribution of 
\be
\chihat=\f{z_1+z_2}{2}
\label{eq:chihat}
\ee
follows directly from Eqs.~(\ref{eq:pdf_sum}) and~(\ref{eq:pdf_func}). 
\begin{widetext}
\be \label{eq:pfield}
\hat{p}_{\rm field}(\chihat) \! =\f{4}{\delta^2}\! \times\!
\begin{cases}
\xi ^2 \chihat & \text{for } 0 \leq \chihat \leq (1-\delta)/2 , \\
 -\xi  (\delta +(\xi +2) \chihat-2 \chihat \log (2 \chihat)-1) & \text{for } (1-\delta)/2 \leq \chihat \leq 1/2 , \\
-\xi  (\delta +\xi  \chihat) & \text{for } 1/2 \le \chihat\le 1-\delta , \\
\chihat \left[\text{Li}_2\left(\frac{1-\delta }{2 \chihat}\right)-\text{Li}_2\left(\frac{2 \chihat+\delta -1}{2 \chihat}\right)\right]-\xi  [\chihat \log (2 (\delta +2 \chihat-1))+\chihat\log\chihat&\\-2 \chihat+1]
+2 (\delta +\chihat-1)+(-\delta -2 \chihat+\chihat \log (2 \chihat)+1) \log (\delta +2 \chihat-1) & \text{for }  1-\delta \le \chihat \le 1-\delta/2 , \\
\chihat \text{Li}_2\left(1-\frac{1}{2 \chihat}\right)-\chihat \text{Li}_2\left(\frac{1}{2 \chihat}\right)\!-2 \chihat\!-(-2 \chihat+\chihat \log (2 \chihat)+1) \log (2 \chihat-1)+2 & \text{for } 1-\delta/2 \leq \chihat \leq 1 , \\
\end{cases}
\ee
\be \label{eq:pfield2}
\hat{p}_{\rm field}(\chihat) \! =\f{4}{\delta^2}\! \times\!
\begin{cases}
\xi ^2 \chihat & \text{for } 0 \leq \chihat \leq (1-\delta)/2 , \\
 -\xi  (\delta +(\xi +2) \chihat-2 \chihat \log (2 \chihat)-1) & \text{for } (1-\delta)/2 \leq \chihat \leq (1-\delta) , \\
\chihat \left[\text{Li}_2\left(\frac{1-\delta }{2 \chihat}\right)- \text{Li}_2\left(\frac{2 \chihat+\delta -1}{2 \chihat}\right)\right]+2 (\delta +\chihat-1)\\-\left(\delta -\chihat \log \left(\frac{2 \chihat}{1-\delta}\right)+2 \chihat-1\right) \log (\delta +2 \chihat-1)+\xi  \chihat \log (2 \chihat) & \text{for } (1-\delta)  \le \chihat\le 1/2 , \\
\chihat \left[\text{Li}_2\left(\frac{1-\delta }{2 \chihat}\right)- \text{Li}_2\left(\frac{2 \chihat+\delta -1}{2 \chihat}\right)\right]+2 (\delta +\chihat-1)\\-\left(\delta -\chihat \log \left(\frac{2 \chihat}{1-\delta}\right)+2 \chihat-1\right) \log (\delta +2 \chihat-1)-\xi  \left(\chihat \log (2 \chihat)-2 \chihat+1\right) & \text{for }  1/2 \le \chihat \le 1-\delta/2 , \\
\chihat \text{Li}_2\left(1-\frac{1}{2 \chihat}\right)-\chihat \text{Li}_2\left(\frac{1}{2 \chihat}\right)\!-2 \chihat\!-(-2 \chihat+\chihat \log (2 \chihat)+1) \log (2 \chihat-1)+2 & \text{for } 1-\delta/2 \leq \chihat \leq 1 , \\
\end{cases}
\ee
\end{widetext}
where Eqs.~(\ref{eq:pfield}) hold for $\delta\le1/2$ ($\theta_{\rm max}=60^\circ$), Eqs.~(\ref{eq:pfield2}) hold for $1/2<\delta\leq 1$, we defined $\xi = \log(1-\delta)$, and
\be
\text{Li}_2(z)=\sum_{k=1}^\infty \f{z^k}{k^2}=\int_{z}^0\f{\ln(1-t)}{t}dt
\ee
is the dilogarithm.

\subsection{Cluster binaries}

\subsubsection{First generation}

For cluster BBHs we assume an isotropic distribution, i.e. we draw $\cos\theta_i$ from a uniform distribution in $[-1,1]$.
First we find the distribution of $z_i=\chihat_{i} \cos\theta_{i}$, which is a product of two uniform distributions:
\bea
p(\cos\theta_i) &= {1}/{2}&, \quad \cos\theta_i  \in [-1,1]\,,\nn\\
p(\chihat_i) &= 1&, \quad  \chihat_i  \in [0,1]\,.
\eea
Using Eq.~(\ref{eq:pdf_prod}) one has
\be
p(z_i)=-\f{1}{2}\log|z_i|, \quad \text{for } z_i\in [-1,1]\,.
\ee
Finally, the distribution of $\chihat$  from Eq.~(\ref{eq:chihat})
can be calculated using Eqs.~(\ref{eq:pdf_sum}) and (\ref{eq:pdf_func}), with the result
\begin{widetext}
\be \label{eq:pcluster}
\hat{p}_{\rm cluster}(\chihat)=
\begin{cases}
 -2|\chihat| \text{Li}_2\left(\frac{2 |\chihat|}{2 |\chihat|-1}\right)-\frac{1}{2} \left(4+\pi ^2\right) |\chihat|-\log (1-2 |\chihat|) \
(-2 |\chihat|+|\chihat|\log (1-2 |\chihat|)+1)+2 & \text{for } 0 <|\chihat|  \leq 1/2 , \\
    |\chihat| \text{Li}_2\left(1-\frac{1}{2 |\chihat|}\right)-|\chihat| \text{Li}_2\left(\frac{1}{2 |\chihat|}\right)-2 \chihat-(-2 \
|\chihat|+|\chihat| \log (2 |\chihat|)+1) \log (2 |\chihat|-1)+2 & \text{for } 1/2 < |\chihat| \leq 1 .
\end{cases}
\ee
\end{widetext}
\subsubsection{2g+1g and 2g+2g mergers}
\label{app:chieff2g}

Here we provide some approximations to the PDFs of 2g mergers. These are not used explicitly in the main body of the paper, but they are useful to understand some of the trends observed in our model. %

Let us start from 2g+1g events. In the small-$\chimax$ limit, the spin of 1g BHs is neglibigle, while 2g remnants will have spins $\chi_f\simeq0.68$, which yields
\be
\chieff \simeq \f{\chi_f \cos\theta_1}{1+q}\,.
\ee
Because $\beta\gg 1$, 2g BH with $m_1<\mmax$ will pair with a 1g BH of similar mass (i.e. $q=1$), resulting in
$\chieff  \simeq 0.34 \cos\theta_1$. Since $\cos\theta_1$ is distributed uniformly in $[-1,\,1]$, the resulting distribution of $\chieff$ is also uniform in $[-0.34, 0.34]$:
\be
p_{2g+1g}(\chieff) \simeq \f{1}{\chi_f} \; \;\;\text{for} \;\; |\chieff|\le \f{\chi_f}{2}\,,
\ee
as seen in the bottom panel of Fig.~\ref{fig:hist}. The equal-mass assumption breaks down for 2g+1g mergers   in the mass gap, which leads to events with $|\chieff| > 0.34$, causing the ``tail'' in the $\chieff$ distribution observed in Fig.~\ref{fig:hist}.   %

For 2g+2g mergers, the equal-mass approximation remains appropriate. One has 
\be
\label{p22app}
\chieff \simeq \f{\chi_f}{2} \,(\cos\theta_1+\cos\theta_2)\,.
\ee 
Both $\cos\theta_i$ are distributed uniformly in $[-1,1]$.  Eq.~(\ref{eq:pdf_sum}) returns a PDF
\be
p_{2g+2g}(\chieff) \simeq \f{1}{\chi_f} \left(1 \!-\! \f{\vert\chieff\rvert}{\chi_f}\right)\; \;\;\text{for} \;\; |\chieff|\le\chi_f\,,
\ee
 in good agreement with Fig.~\ref{fig:hist}.

With the above distributions, we can also provide an analytical approximation for the probability that an event lies in the spin gap:
\bea
p_{\rm 2g+1g}(|\chieff|>\chimax) &=& 1- \f{2 \chimax}{\chi_f}\, ,\label{p21spingapapp}\\
p_{\rm 2g+2g}(|\chieff|>\chimax) &=& \left(1- \f{\chimax}{\chi_f}\right)^2\, .\nn\\
\eea

Suppose that  a fraction $f_{\rm 2g+2g}$ of all 2g events are assigned to 2g+2g (cf. Sec.~\ref{sec:2gvs1g}), so that
\be
p_{2g}(\chieff)= (1-f_{\rm 2g+2g})\, p_{\rm 2g+1g}(\chieff)+ f_{\rm 2g+2g} \, p_{\rm 2g+2g}(\chieff)\, .
\ee
The probability of having a spin-gap event is given by
\be
p_{\rm 2g}(|\chieff|>\chimax) = 1-\frac{2 \chimax}{\chi_f}+f_{\rm 2g+2g}\left(\frac{\chimax}{\chi_f}\right)^2\, .
\ee
The term of order $\mathcal{O}(\chimax^2)$ can be neglected for small values of $\chimax$. Furthermore, let us note that $f_{\rm 2g+2g}$ is a monotonically decreasing function of $\chimax$, with maximum $\simeq 0.25$. 
We can thus approximate the spin-gap efficiency as being independent of $f_{\rm 2g+2g}$:
\be
\lspin(\chimax) \simeq \left(1-\frac{2 \chimax}{\chi_f}\right) \lambda_{\rm 2g}(\chimax)\,, ,
\ee
where $\lambda_{\rm 2g}(\chimax)$ is the efficiency of producing 2g events, i.e. the fraction of all events that are either 2g+2g or 1g+2g.

\bibliography{clusters}

\end{document}